\newcommand\papertitle{Frequency dependence of the thermal dust $E/B$ ratio and $EB$ correlation: Insights from the spin-moment expansion}

\documentclass{aa}

\makeatletter
\renewcommand*\aa@pageof{, page \thepage{} of \pageref*{LastPage}}
\makeatother
\usepackage{bbm}
\usepackage{graphicx}
\usepackage{txfonts}
\usepackage{url}
\usepackage{natbib}
\bibpunct{(}{)}{;}{a}{}{,} 
\usepackage{color}
\usepackage{colortbl} 
\usepackage[dvipsnames]{xcolor}
\usepackage{enumitem}
\usepackage{tipa}
\usepackage{booktabs}
\usepackage[switch]{lineno}

\usepackage[breaklinks, colorlinks, citecolor=blue]{hyperref}
\usepackage{subfigure}
\usepackage{epsfig}
\usepackage{multirow}
\usepackage{array}

\usepackage{url}
\usepackage{psfrag}
\usepackage[T1]{fontenc}
\usepackage{rotating}
\usepackage{soul}
\usepackage{wrapfig}
\usepackage{tikz}

\makeatletter
\def\env@cases{
  \let\@ifnextchar\new@ifnextchar
  \left\lbrace
  \def\arraystretch{1.2}
  \array{l@{}l@{}}
}
\makeatother

\definecolor{mygreen}{RGB}{104,198,107}
\definecolor{myred}{RGB}{252,137,125}
\definecolor{myyellow}{RGB}{252,225,126}
\definecolor{mygrey}{RGB}{215,215,215}

\def\fracnu{\left(\frac{\nu}{\nu_0}\right)}
\def\lnnu{\ln\fracnu}
\def\psinu{\psi(\nu)}
\def\i{\mathbbm{i}}
\def\d{\rm d}
\def\blackbody{B^{\rm Pl}}

\def\planck{\textit{Planck}}

\def\dzero{{\tt d0}}
\def\done{{\tt d1}}
\def\deight{{\tt d8}}

\def\dten{{\tt d10}}
\def\dtwelve{{\tt d12}}

\newcommand{\expf}[1]{{\rm e}^{#1}}

\newcommand{\spinscalmom}{\mathbb{W}}
\newcommand{\EBangle}{\vartheta}

\def\barA{\bar{A}}

\def\spinpol{\mathcal{P}}
\def\hatpol{\varepsilon^{P}}

\def\DLEE{\mathcal{D}_\ell^{EE}}
\def\DLBB{\mathcal{D}_\ell^{BB}}
\def\DLEB{\mathcal{D}_\ell^{EB}}
\def\psibg{\mathcal{D}_\ell^{EB}}
\def\psifl{\psi^{\rm fl}}
\def\psibg{\psi^{\rm bg}}

\begin{document}

\title{\papertitle}

\offprints{\url{leo.vacher@irap.omp.eu}}
\authorrunning{Vacher et al.}
\titlerunning{Frequency dependence of the $E-B$ ratio and $EB$ correlation.}

\author{L. Vacher\inst{\ref{inst1}} 
\and J. Aumont\inst{\ref{inst1}}
\and F. Boulanger\inst{\ref{inst2}}
\and L. Montier\inst{\ref{inst1}}
\and V. Guillet\inst{\ref{inst3},\ref{inst4}}
\and A. Ritacco\inst{\ref{inaf},\ref{inst2}}
\and J. Chluba \inst{\ref{Manch}}
}
\institute{
IRAP, Universit\'e de Toulouse, CNRS, CNES, UPS, 9 Av. du Colonel Roche, 31400 Toulouse, France\label{inst1}
\and
Laboratoire de Physique de l’Ecole normale supérieure, ENS, Université PSL, CNRS, Sorbonne Université
Université Paris-Diderot, Sorbonne Paris Cité, 24 rue Lhomond
75005 Paris, France\label{inst2}
\and
Institut d'Astrophysique Spatiale, CNRS, Universit\'{e} Paris-Saclay, CNRS, Rue Jean-Dominique Cassini B\^{a}t. 121, 91405 Orsay, France\label{inst3}\goodbreak
\and
Laboratoire Univers et Particules de Montpellier, Universit\'e de
Montpellier, CNRS/IN2P3, CC 72, Place Eug\'ene Bataillon, 34095
Montpellier Cedex 5, France
\label{inst4}
\and 
INAF-Osservatorio Astronomico di Cagliari, Via della Scienza 5, 09047 Selargius, Italy \label{inaf}
\and 
Jodrell Bank Centre for Astrophysics, Alan Turing Building, University of Manchester, Oxford Rd M13 9PL, Manchester, United Kingdom \label{Manch}
}

\abstract{The change of physical conditions across the turbulent and magnetized interstellar medium induces a 3D spatial variation of the properties of Galactic polarized emission. The observed signal results from the averaging of different spectral energy distributions (SEDs) and polarization angles along and between lines of sight. As a consequence, the total Stokes parameters $Q$ and $U$ will have different {frequency dependencies, both departing from the canonical emission law}, so that the polarization angle becomes frequency dependent. In the present work, we show how this phenomenon similarly induces a different, distorted SED for the three polarized angular power spectra $\DLEE$, $\DLBB$, and $\DLEB$, implying a variation of the $\DLEE/\DLBB$ ratio with frequency.
We demonstrate how the previously introduced {"spin-moment"} formalism provides a natural framework to grasp these effects and enables us to derive analytical predictions for the spectral behaviors of the polarized spectra, focusing here on the example of thermal dust polarized emission. After a quantitative discussion based on a model combining emission from a filament with its background, we further reveal that the spectral complexity implemented in the dust models commonly used by the cosmic microwave background (CMB) community includes different distortions for the three polarized power-spectra. This new understanding is crucial for CMB component separation, in which extreme accuracy is required for the modeling of the dust signal to allow for the search of the primordial imprints of inflation or cosmic birefringence. For the latter, as long as the dust $EB$ signal is not measured accurately, great caution is required regarding the assumptions made to model its spectral behavior, as it may not be inferred from the other dust angular power spectra.}


\maketitle

\section{Introduction}
Understanding Galactic foregrounds is a critical challenge for the success of cosmic microwave background (CMB) experiments searching for primordial $B$-modes leftover by inflationary gravitational waves  \citep[see e.g.,][]{Kamionkowski16} and  signatures of cosmic birefringence \citep[see e.g.,][]{PIPXLIX_birefringence,Diego-Palazuelos2022a}. In these quests, both the structure on the sky 
and the frequency-dependence of the foreground signal need to be modeled.

The two-point statistical properties of a polarized signal are described by angular auto- and cross-power spectra, hereafter simply written as $XY$, where $X$ and $Y$ refer to $E-$ and $B-$mode polarization or the total intensity $T$. Thermal dust polarization is the main polarized foreground at frequencies above approximately 70 GHz \citep{krachmalnicoff1}.  
%
Based on observations of the \planck{} satellite at 353\,GHz, dust power spectra in polarization have been found to be well fitted by power laws in $\ell$ of similar indices with an $EE/BB$ power ratio of about two. A positive $TE$  and a weaker parity violating $TB$ signal have been significantly detected using the same dataset \citep{Weiland2020,PlanckDust,PlanckDust2}. The dust $EB$ signal, however, remains compatible with zero at \planck\ sensitivity \citep{PlanckDust2}. 

The $EE/BB$ asymmetry and $TE$ correlation relate to the anisotropic structure of the magnetized interstellar medium with filamentary structures in total intensity preferentially aligned with the Galactic magnetic field \citep{Clark15,PlanckXXXVIII}. { Statistical properties of dust polarization have been discussed on theoretical grounds as signatures of magnetized interstellar turbulence \citep{Caldwell17,Kandel18,Bracco19}. Various empirical and phenomenological models have been proposed \citep{Ghosh17,Clark19,Huffenberger2020,Hervias22,Konstantinou22}. Within a phenomenological framework}, a coherent misalignment between filamentary dust structures and the magnetic field can account for the dust $TB$ signal and should also imply a positive $EB$ \citep{Clark2021,Cukierman2022}. The possibility of a nonzero dust $EB$ signal is at the heart of recent analyses of \planck\ data that seek to detect cosmic birefringence because it 
complicates attempts to measure a CMB-$EB$ correlation \citep{Minami19,Diego-Palazuelos2022a,Eskilt22,Diego-Palazuelos2022b}.

While evaluating the amplitudes of the foreground $EE/BB$ ratio and $EB$ correlation are central subjects in the literature, their frequency dependence is rarely discussed. In the present work, we intend to open this discussion. Given  
the level of accuracy targeted by future experiments \citep[e.g.,][]{Ptep,CMBS4}, the frequency dependence of dust polarization is a critical issue for CMB component separation, which
relates CMB experiments to the modeling of dust emission \citep{Guillet18,Hensley22}.
The \planck\ data have also been crucial in building our current understanding of this topic \citep{PIPXXII,PlanckDust}. The mean spectral energy distribution (SED), derived from $EE$ and $BB$ power spectra, is found to be well fitted by a single modified blackbody (MBB) law \citep{PlanckDust2}, but this empirical law does not fully characterize the frequency dependence of dust polarization.
Indeed, integration along the line of sight and within the beam 
of multiple polarized signals with different spectral parameters and polarization angles must induce departures from the MBB coupled to variations of the total polarization angle with frequency \citep{tassis,PIPL,Vacher2022b}.\footnote{{These behaviors are sometimes referred to as {"frequency decorrelation"} 
(see e.g., \cite{pelgrims2021}).}}This effect has been detected in \planck{} data by \citet{pelgrims2021} for a discrete set of lines of sight selected based on H~I data. 
From a power spectra analysis of {\it Planck} polarization maps, \citet{Ritacco22} revealed an unanticipated impact of the polarization angles' frequency dependence on the decomposition of the dust polarized emission into $E$- and $B$-modes.
Their result emphasizes the need to account for variations of polarization angles in order to model the dust foreground to the CMB. 


\begin{figure}[t]
    \centering
    \includegraphics[scale=0.38]{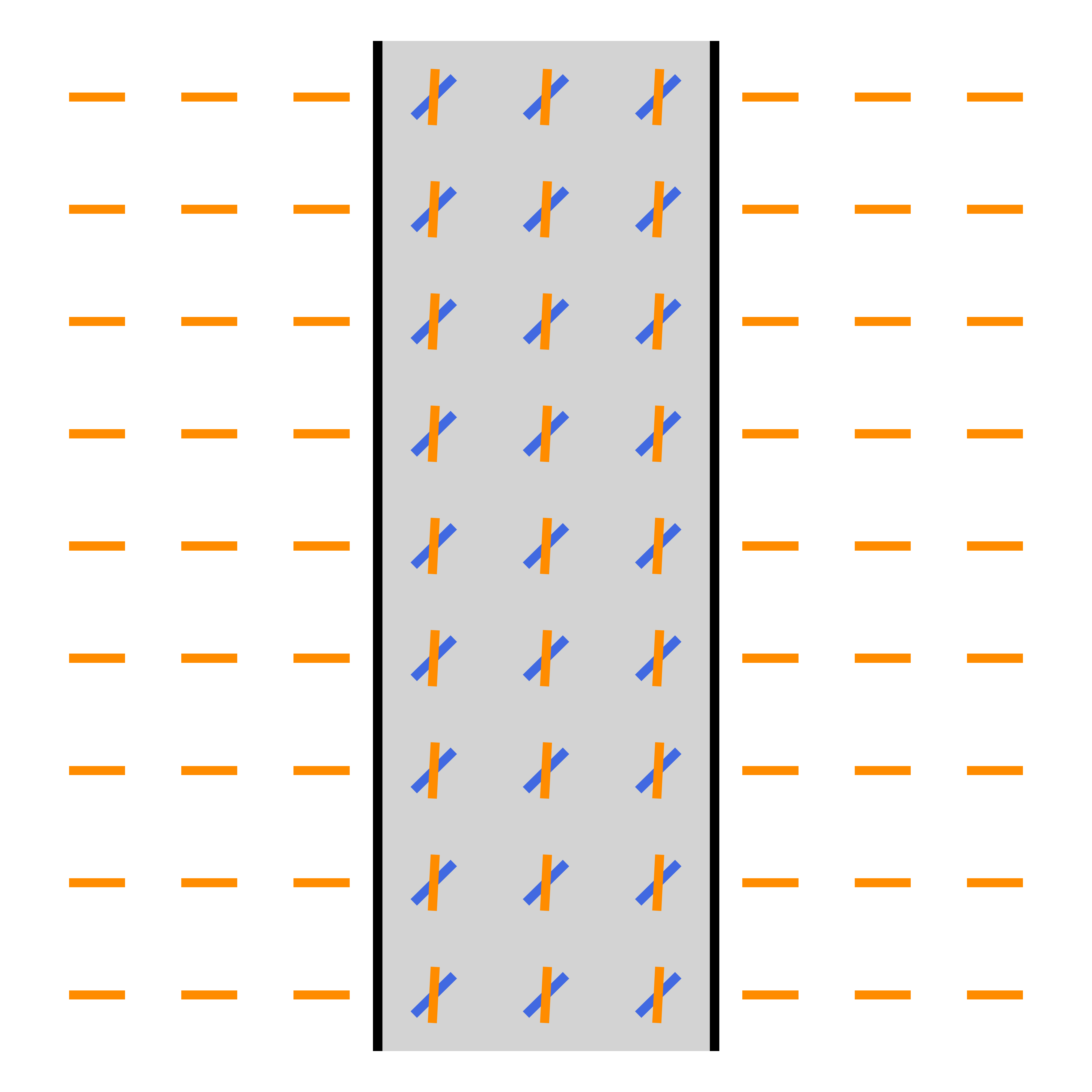}
    \caption{ Diagram of the toy model composed of an infinite filament (grey) over a background (white). The orientation of the $\psi(\nu)$ field is represented with color bars at two different frequencies: $\nu_1$ (orange) and $\nu_2$  (blue). 
    }
    \label{fig:toy-model}
\end{figure}

The moment expansion formalism, introduced in intensity by \citet{Chluba2017}, proposes to treat the distortions from averaging over different emission points by Taylor expanding the canonical SED -- the MBB for dust polarization -- with respect to its spectral parameters. This framework has proven to be a powerful tool for component separation and Galactic physics \citep{Remazeilles_etal_2016,RemazeillesmomentsILC,MILC,Sponseller2022}. When carried out at the power-spectrum level, as in \citet{Mangilli}, the expansion can be applied directly to the $B$-mode signal as an intensity \citep{Azzoni2020,Vacher2022a,Ritacco22}. A recent generalization of this formalism to polarization in \citet{Vacher2022b}, that is, the "spin-moment" expansion, provides a natural framework to treat for the spectral dependence of the polarization angle. In the present paper, we provide the missing links between the spin-moment maps and the treatment of $E$- and $B$-modes. We hence connect the statistical studies of the sky maps with the modeling of the frequency dependence of dust polarization angular power spectra, enabling the prediction of new consequences unique to polarization. In particular, we discuss how and when the assumption of a common SED for $EE$ and $BB$ (and $EB$) stops being valid.

This article is organized as follows. In Sec.~\ref{sec:eb_presentation}, we introduce the specificity of the polarized signal and explain qualitatively why we expect the frequency dependence of $EE$, $BB$, and $EB$ to differ. In Sec.~\ref{sec:spin-moments}, we establish how the formalism of the spin-moments can naturally describe these effects and give an analytical expression for their frequency dependence. In Sec.~\ref{sec:toy-model}, we illustrate the derived formalism on a filament model. Then, in Sec.~\ref{sec:Pysm}, we show that the frequency dependence of the $EE/BB$ ratio and the nontrivial dependence of $EB$ are already present in dust models extensively used by the CMB community, which 
 brings us to stress the need for caution when inferring the spectral properties of the dust $EB$ signal from other angular power spectra.  Finally, we present our conclusions in Sec.~\ref{sec:conclusion}.

\section{Combination of polarized signals}
\label{sec:eb_presentation}
\subsection{Mixing of polarized signals in the {\it Q}-{\it U} plane \label{sec:QUmix}}

The frequency-dependent linear polarization\footnote{We do not discuss the circular polarization quantified by $V_\nu$ in this work.} of a signal is described by a {"polarization spinor"} $\spinpol_\nu$.
\footnote{Note that here, as in \cite{Vacher2022b}, we use the word "spinor" in a  general sense which must be understood as "spin-$s$ field on the sphere".}.
It is a complex-valued object that can be expressed both in Cartesian and exponential form as
\begin{equation}
    \mathcal{P}_\nu \equiv Q_\nu + \i U_\nu = P_\nu e^{2\i\psinu},
    \label{eq:spinor}
\end{equation}
where $\i^2=-1$ is the imaginary unit. The spinor components ($Q_\nu$,$U_\nu$) are two of the Stokes parameters, and they are proportional to the difference of intensities in two orthogonal directions rotated from one  another by an angle of $45^\circ$. Reportedly, $\spinpol_\nu$ is a spin-2 object due to its transformation properties under rotations. It rotates in the complex plane by an angle $-2\theta$ under a right-handed rotation around the line of sight by an angle $\theta$ of the coordinates in which the Stokes parameters are defined.

The modulus of the spinor in Eq.~\eqref{eq:spinor} is the linear {"polarized intensity"}
\begin{equation}
    P_\nu = \sqrt{Q_\nu^2 + U_\nu^2},
\end{equation}
encoding the SED of the total polarized signal. Depending on the physics of the emission, $P_\nu$ can be a function of various parameters, $\vec{p}$, called the {"spectral parameters."}

Both the total intensity and the polarized intensity of the thermal dust grains at a given point of the Galaxy are expected to follow an MBB law \citep[see e.g.,][]{Planck2015XX}
\begin{equation}
    P^{\rm MBB}_\nu = A\fracnu^{\beta}\blackbody_\nu(T) = A\hatpol_\nu(\beta,T) ,
\end{equation}
with the corresponding spectral parameters being $\vec{p}=\{\beta,T\}$, where  $\beta$ is the spectral index, related to dust grain properties, and $T$ is the temperature associated to the blackbody law $\blackbody_\nu$. The arbitrary reference frequency $\nu_0$ is used for normalization, and the dust polarization amplitude  $A= p_0\tau \cos^2(\gamma)$ can be expressed in terms of the intrinsic degree of polarization $p_0$, the dust {optical depth $\tau$ evaluated at $\nu_0$}, and the angle between the Galactic magnetic field and the plane of the sky $\gamma$ (see e.g., \citet{Planck2015XX}). The polarized emissivity function $\hatpol_\nu(\beta,T)$ encodes all the spectral dependence of the MBB law.\footnote{For comparison, in \cite{Vacher2022b}, $\hatpol_\nu$ was as written $\hat{P}^{\rm mbb}_\nu$ {and $\psinu$ was as written $\gamma_\nu$.}}

The spinor phase in Eq.~\eqref{eq:spinor} is the so-called polarization angle\footnote{The polarization angle is defined here in the IAU convention (counted positively from Galactic north toward Galactic east), provided that both $Q$ and $U$ also respect this convention. If, however, the healpix convention is used (as is the case for \planck{} data, for example), $U$ must be replaced by $-U$ to recover the IAU defined convention for $\psi.$ (See e.g., \cite{Planck18_XII}).}
\begin{equation}
    \psi = \frac{1}{2}\arctan\left( \frac{U_\nu}{Q_\nu} \right).
\label{eq:psi}
\end{equation}
This definition explicitly shows that if $Q_\nu$ and $U_\nu$ have the same SED, $\psi$ is a constant number defining a single orientation at all frequencies. This assumption is usually made when locally modeling the dust signal, as $\psi$ is orthogonal to the local orientation of the Galactic magnetic field,\footnote{The two angles $\psi$ and $\gamma$ together quantify the 3D orientation of the Galactic magnetic field and as such are not independent quantities. In the remainder of this work, however, we consider the case of a constant $\gamma,$ and we do not further discuss the statistical dependence of the two angles.} which is well motivated by the behavior of the elongated dust grains. If however $Q_\nu$ and $U_\nu$ behave differently, the polarization orientation defined by $\psi$ will change with frequency so that $\psi \to \psinu$.

In observational conditions, the mixing of multiple signals coming from emission points with different physical conditions are unavoidable along the line of sight $\vec{n},$  between the lines of sight, inside the instrumental beam, or over patches of the sky when using spherical harmonic transformations. As discussed in \citet{Chluba2017}, this mixing will induce departures from the canonical SED of the total intensity, known as "SED distortions," that can be properly modeled using a Taylor expansion of the signal with respect to the spectral parameters themselves. This expansion is known as "moment expansion". The mixing also has unique consequences in polarization. In the rest of this work, we refer to the combination of individual polarized signals with different spectral parameters and polarization angles as the "polarized mixing." The resulting spinor obtained through polarized mixing will inherit both the SED distortions and a spectral dependence of the polarization angle, which can be modeled using a complex moment expansion of $\spinpol_\nu$ \citep{Vacher2022b}. In principle, it is even possible to predict the value of $\psinu$ from the distribution of polarization angles and spectral parameters. For example, considering a sum of MBBs with different polarization angles and spectral indices, one expects at first order that
\begin{equation}
    \psinu \simeq \psi(\nu_0) + \frac{{\rm Im}(\mathbf{{\Delta \beta}})}{2}\lnnu,
\label{eq:psi_nu_lnnu}
\end{equation}
with $\mathbf{{\Delta \beta}} \in \mathbb{C}$ as the complex spectral index correction,
\begin{equation}
    \mathbf{\Delta \beta} 
    = \frac{\langle A \expf{2\i \psi} (\beta - \bar{\beta})\rangle}{ \langle A \expf{2\i \psi} \rangle},
\label{eq:complex-beta}
\end{equation}
and $\langle \dots \rangle$ indicating sums or integrals along the line of sight or in the beam of the $A$, $\beta$, and $\psi$ distributions. The value represented by  $\bar{\beta}$ is the pivot spectral index around which the expansion is performed.

\subsection{Mixing of polarized signals in the {\it E}-{\it B} plane}
\label{sec:EBmix}

In analogy with electromagnetism, the $E-$ and $B-$modes are scalar and pseudo-scalar fields respectively quantifying  the existence of curl-free and divergence-free patterns of the $\psi$ field over the sky. As such, the $E$ and $B$ fields are nonlocal quantities equivalent to convolutions of the $Q$ and $U$ fields around each point of the sky.
From a generalization of the Helmholtz decomposition theorem, a linearly polarized signal can be fully split into $E$ and $B$ components. The $E$- and $B$-modes of the polarized dust emission are frequency dependent quantities. When all the emission points share the same spectral parameters, $E_\nu$ and $B_\nu$ will have the same SED. 
In this case, the $EE/BB$ ratio is constant with frequency, and the $EB$ correlation has the same SED as $EE$ and $BB$.
As we show, polarized mixing imposes a different SED for $E_\nu$ and $B_\nu$ (as $Q_\nu$ and $U_\nu$), and in such cases, the three angular power spectra should inherit a different distorted SED, while the $EE/BB$ ratio becomes frequency dependent.

To illustrate this effect, we start with a very simple example inspired from \citet{Zaldarriaga2001} and \citet{PlanckXXXVIII,Planck2016_XXXIII}. As in Fig.~\ref{fig:toy-model}, we consider an infinite filament\footnote{The situation would be different for a finite filament. On this point, see the discussions in \citet{Rotti2019}.} in front of a polarized background. If the background and the filament have different polarization angles and spectral parameters, polarized mixing will unavoidably occur from their superposition, and the resulting $\psi$ field over the filament will rotate with frequency.
As sketched out with  Fig.~\ref{fig:toy-model}, one would imagine that at a given frequency, $\nu_1$, the configuration is such that $\psi(\nu_1)$ in the filament is perpendicular to $\psi(\nu_1)$ in the background. Similar to the case of \cite{Zaldarriaga2001}, such a pattern is curl-free, and one expects to find $B_{\nu_1}=0$ and that the signal can be purely described by $E$-modes.
Due to the spectral rotation, at any other frequency ($\nu_2$), $\psi(\nu_2)$ will be rotated uniformly over the filament, necessarily leading to a different configuration in which $B_{\nu_2}\neq 0$. It is then clear that the signal's $EE/BB$ ratio must be a frequency dependent quantity. For this to be possible, one would hence conclude that $E_\nu$ and $B_\nu$ do not share the same SED anymore and neither do $EE$, $BB,$ and $EB$.

\section{Insights from the spin-moments}
\label{sec:spin-moments}

In this section, we explore how to quantify and model the frequency dependence of the $EE/BB$ ratio and the distortions of the $EB$ correlation predicted in Sec.~\ref{sec:EBmix}. As we reveal, the spin-moment expansion formalism provides a natural framework to do so. To keep our discussion as simple as possible, we consider only the case of averaging MBB with different values of $A$, $\beta$, and $\psi$ and we display  expansions only up to first order. All the following derivations can however be straightforwardly  generalized at any order and used to consider variations of temperatures. All our conclusions about the behavior of the spectra would remain true for any choice of SED (e.g., for the synchrotron signal).

\subsection{Q/U spin-moment expansion}

As presented in Sec.~\ref{sec:QUmix}, we consider that the dust grain polarized signal is given locally by the spinor $\spinpol_\nu$ in every point of the Galaxy, with $P_\nu = |\spinpol_\nu|$ as an MBB and $\psi$ as a frequency independent quantity. The average spinor over different emission points centered on a given line of sight $\vec{n}$ is given by the spin-moment expansion around an arbitrary pivot value $\bar{\beta}$\footnote{The choice of the weighted average $\bar{\beta} = \langle A\beta \rangle / \langle A \rangle$, which cancels the first order moment in intensity, is expected to be best suited for convergence.}
for the spectral index as
\begin{equation}
    \left<\mathcal{P}_\nu(\vec{n})\right>    = 
    \hatpol_\nu(\bar{\beta},\bar{T})
    \left(\mathcal{W}_{0}
    \, + \mathcal{W}_1^{\beta} \lnnu + \dots \right).
\label{eq:spin-moment-expansion}
\end{equation}
The spin-moments $\mathcal{W}_{k}^\beta$ of order $k$ associated to the spectral index can be estimated directly from the distribution of $\psi$, $A$, and $\beta$ as\footnote{For simplicity, we chose a different choice of normalization for the spin-moments than \citet{Vacher2022b}, simplifying the $\barA$ of the derivatives with the ones in the original definition of the spin-moments.}
\begin{equation}
\label{eq:spin-moments}
   \mathcal{W}_{k}^{\beta} 
    =\left<A\,\expf{2\i\psi} (\beta-\bar{\beta})^{i}\right>,
\end{equation}
where $\mathcal{W}_{0}=\langle A e^{2\i\psi}\rangle$ plays the role of a total complex amplitude that satisfies  $|\mathcal{W}_{0}| \leq \langle A\rangle$. Purely geometrical phenomenon  can greatly decrease the value of $\mathcal{W}_0$. For example, the cancellation of the phases known as "depolarization" $(\langle e^{2\i\psi} \rangle \simeq 0$),  or a significant inclination of the Galactic magnetic field toward the plane of the sky ($\cos(\gamma) \simeq 0$), would both lead to $\mathcal{W}_0 \simeq 0$. The condition $|\mathcal{W}_0|\gg |\mathcal{W}^\beta_k|$ defines the {"perturbative regime"} 
such that one can consider the total signal as a perturbed MBB. However, as discussed in \citet{Vacher2022b}, one would expect the existence of configurations where the canceling effects are strong enough such that the total signal is mostly or fully given by its moments and thus looses its MBB behavior. In general, the phase weighting\footnote{While the two angles $\psi$ and $\gamma$ have a common geometrical origin in the Galactic magnetic field, they play a very different role here. The angle $\psi$ is a complex phase allowing for "interference"-type cancellation of the moment terms, while $\cos^2(\gamma)$ (as $p_0$) plays the role of a real and positive weight.} in Eq.~\eqref{eq:spin-moments}, which is unique to polarization, breaks the expected hierarchy between the moments, and one would not be able to ascertain that $\mathcal{W}^\beta_k > \mathcal{W}^\beta_m$ solely because $k > m$.

The expansion can equivalently be split into $Q$ and $U$ coordinates as
\footnote{By treating the expansions of $Q$ and $U$ independently (instead of considering them together in $\spinpol_\nu$), one would loose the information on their correlation.}
\begin{align}
\mathcal{W}_{k,Q}^{\beta} &= {\rm Re}(\mathcal{W}_{k}^{\beta}),\\
\mathcal{W}_{k,U}^{\beta} &= {\rm Im}(\mathcal{W}_{k}^{\beta}).
\end{align}
Different moments in $Q$ and $U$, expected in the general case, will necessarily imply a frequency dependence of the polarization angle $\psi \to \psinu$ from the definition given in Eq.~\eqref{eq:psi}. In the perturbative regime, the first order $\mathcal{W}^\beta_1/\mathcal{W}_0$ can be interpreted as a complex correction to the pivot spectral index, leading to 
\begin{equation}
    \psinu \simeq \psi(\nu_0) + \frac{1}{2}{\rm Im}\left(\frac{\mathcal{W}_1^\beta}{\mathcal{W}_0}\right)\lnnu,
\label{eq:psi_nu_lnnu2}
\end{equation}
and giving back the result mentioned in Eq.~\eqref{eq:psi_nu_lnnu}. We note again that both variations of the polarization angles and the spectral indices are required for the second term to be nonzero.

\subsection{\texorpdfstring{$E_\nu$}{Enu} and \texorpdfstring{$B_\nu$}{Bnu} in map space}
As for $Q$ and $U$ in Eq.~\eqref{eq:spinor}, the $E$ and $B$ fields can be grouped in a single complex scalar field \citep{Zaldarriaga1997}:
\begin{equation}
    \mathcal{S}_\nu \equiv E_\nu + \i B_\nu = S_\nu\expf{\i \EBangle(\nu)},
\end{equation}
of modulus and phase (which we refer to as the $E$-$B$ angle)
\begin{align}
    S_\nu &= \sqrt{E_\nu^2 + B_\nu^2},\\
    \EBangle &= \arctan \left( \frac{B_\nu}{E_\nu} \right).
\end{align}
The frequency dependence of the $EE/BB$ ratio discussed in Sec.~\ref{sec:EBmix} would then  translate itself into the rotation of $\mathcal{S}_\nu$ in the complex plane, and the $E$-$B$ angle would become frequency dependent, that is, $\EBangle\to\EBangle(\nu)$.

A straightforward way to obtain the $\mathcal{S}_\nu$ field from the previously introduced polarization spinor field $\spinpol_\nu$ is to use the spin-raising operator $\bar{\eth}$ -- the conjugate of the spin-lowering operator $\eth$ -- as 
\begin{equation}
    \mathcal{S}_\nu(\vec{n}) = -\bar{\eth}^2 \mathcal{P}_\nu(\vec{n}),
\end{equation}
where $\eth$ is acting both as an angular momentum ladder operator and as a covariant derivative on the sphere (see, e.g., \citet{Goldberg1967,Rotti2019}). As such, it contains derivatives with respect to the spherical coordinates $\theta$ and $\varphi$. More technically, for a spin-s field $\eta$ on the sphere,
\begin{equation}
\bar{\eth} \eta  = -(\sin\theta)^s\left[\partial_\theta - \i \sin(\theta)^{-1} \partial_{\varphi}\right]\left[\sin(\theta)^{-s} \eta\right]. 
\end{equation}
This operator mixes the real and imaginary parts of $\eta$ in a nontrivial way. It is however linear and does not act on the spectral dependence, and as a result, the moment expansion will keep the same structure
\begin{equation}
\label{eq:s-moment}
    \left<\mathcal{S}_\nu(\vec{n})\right>    = 
    \hatpol_\nu(\bar{\beta},\bar{T})
    \left(\spinscalmom_{0}
    \, + \spinscalmom_1^{\beta} \lnnu + \dots \right),
\end{equation}
%
where the scalar moment maps are extracted from the spin-moments as
\begin{equation}
    \spinscalmom_{k}^{\beta} = -\bar{\eth}^2 (\mathcal{W}_{k}^{\beta}).
\end{equation}
As in \citet{Remazeilles_etal_2016,RemazeillesmomentsILC}, one can split the $E$ and $B$ expansions in order to treat them as two scalar spin-0 fields\footnote{The $(Q,U)$ and $(E,B)$ expansions still differ from two independent intensity expansions, as they can take negative values.} (thus loosing their correlation)
\begin{align}
\spinscalmom_{k,E}^{\beta} &= {\rm Re}(\spinscalmom_{k}^{\beta}) = -{\rm Re}(\bar{\eth}^2 (\mathcal{W}_{k}^{\beta})),\\
\spinscalmom_{k,B}^{\beta} &= {\rm Im}(\spinscalmom_{k}^{\beta})=-{\rm Im}(\bar{\eth}^2 (\mathcal{W}_{k}^{\beta})),
\end{align}
where $\eth$ induces a mixing of the $Q$ and $U$ moments into $E$ and $B$ such that different moments for $Q$ and $U$  necessarily imply different moments for $E$ and $B$. 

A clear way to make the action of $\bar{\eth}$ explicit is to consider the flat sky approximation, for which $\bar{\eth} = \partial_x + \i \partial_y$, allowing the expression of the $E$ and $B$ moments map in terms of the two $Q$ and $U$ spin-moment maps as
\begin{align}
\begin{pmatrix}\spinscalmom_{k,E}^{\beta}\\\spinscalmom_{k,B}^{\beta}\end{pmatrix}_{\rm flat} = \begin{pmatrix}\partial_y^2 - \partial_x^2&2\partial_x\partial_y\\-2\partial_x\partial_y&\partial_y^2 -\partial_x^2\end{pmatrix} \begin{pmatrix}\mathcal{W}_{k,Q}^{\beta}\\\mathcal{W}_{k,U}^{\beta}\end{pmatrix}_{\rm flat}.
\end{align}
$E$ and $B$ moments are thus expressed as different linear combinations of second order spatial derivatives of the $Q$ and $U$ spin-moments.

{Whenever a spectral rotation, $\psi \to \psinu$, is induced by polarized mixing, an equivalent rotation,  $\EBangle\to \EBangle(\nu)$, must occur in the $E/B$ plane,} introducing the $EE/BB$ frequency dependence discussed in Sec.~\ref{sec:EBmix}. This rotation can be similarly modeled at first order as
\begin{equation}
    \EBangle(\nu) \simeq \EBangle(\nu_0) + {\rm Im}\left(\frac{\spinscalmom_{1}^{\beta}}{\spinscalmom_{0}}\right)\lnnu.
\end{equation}

\subsection{\texorpdfstring{$E_\nu$}{Enu} and \texorpdfstring{$B_\nu$}{Bnu} in harmonic space}

The spherical harmonic transformation of a spin $s$ field $X$,\footnote{ ${_s}{Y_{\ell m}} \propto (\eth)^s Y_{\ell m}$ ($s>0$) and $\propto (\bar{\eth})^s Y_{\ell m}$ ($s<0$) are the spin-weighted spherical harmonics, defined only for $\ell \geq |s|$.} that is,
\begin{equation}
    (X)_{\ell m} \equiv \int X(\vec{n}) {_s}{Y^*_{\ell m}}(\vec{n})\d^2\vec{n},
\label{eq:Xlm}
\end{equation}
creates additional mixing over angular scales $\ell$. As such, it is expected to increase the moment amplitudes. Consequently, averaging over patches of the sky with a different $\beta$ in each pixel but with no variation along the line of sight is still expected to create SED distortions (see e.g., \cite{Vacher2022a}), and it is expected to make the $EE/BB$ ratio frequency dependent and distort the $EB$ correlation.
Since the transformation is linear, one can simply derive for $\ell \geq 2$\footnote{In order to recover the standard $(E)_{\ell,m}$ and $(B)_{\ell,m}$ as usually defined in cosmology, an extra normalization factor of  $[(\ell-2)!/(\ell+2)!]^{1/2}$ must be added when applying the transformation given by Eq.~\ref{eq:Xlm}, leaving our discussion unchanged.}
\begin{equation}
    (\mathcal{S}_\nu)_{\ell m} = (E_{\nu})_{\ell m} + \i (B_{\nu})_{\ell m}.
\end{equation}
%
With Eq.~\eqref{eq:s-moment}, the expansion becomes
\begin{equation}
    (\mathcal{S}_\nu)_{\ell m}    = 
    \hatpol_\nu(\bar{\beta},\bar{T})    
    \left((\spinscalmom_{0})_{\ell m}
    \, + (\spinscalmom^{\beta}_{1})_{\ell m} \lnnu + \dots \right).
\end{equation}
We note, however, that the pivot $\bar{\beta}$ maximizing the convergence of the expansion in harmonic space might be different than in real space. Hence, one can split the expansion of $E$ and $B$ separately as
\begin{align}
    (\spinscalmom^{\beta}_{k})^{E}_{\ell  m} &= {\rm Re}\left((\spinscalmom^{\beta}_{k})_{\ell m}\right)= \frac{(\spinscalmom_{k}^{\beta})_{\ell m}  + (\spinscalmom_{k}^{\beta})_{\ell m}^* }{2},\\
    (\spinscalmom^{\beta}_{k})^{B}_{\ell m} &= {\rm Im}\left((\spinscalmom^{\beta}_{k})_{\ell m}\right)= \frac{(\spinscalmom_{k}^{\beta})_{\ell m}  - (\spinscalmom_{k}^{\beta})^*_{\ell m} }{2\i}.
\end{align}

\subsection{Power spectra \label{sec:EE-BB}}

\subsubsection{The \texorpdfstring{$EE$}{EE} and \texorpdfstring{$BB$}{BB} power spectra}
When carrying the moment expansion at the power-spectrum level, we found expressions comparable to the ones introduced in intensity by \citet{Mangilli} and applied to $B$-modes in \citet{Azzoni2020,Vacher2022a}. The cross angular power spectra  $\mathcal{D}_\ell^{XX'}$ of two fields $X$ and $X'$ are defined as
\begin{equation}
 \mathcal{D}_\ell^{XX'} = \frac{\ell(\ell+1)}{2\pi}\sum_{m=-\ell}^\ell X_{\ell m} ({X'}_{\ell m})^*,
\end{equation}
with $X,X'\in\{E,B,\spinscalmom_{0,E},\spinscalmom_{0,B},\spinscalmom_{k,E}^\beta,\spinscalmom_{k,B}^\beta\}$. For the $EE$ and $BB$ power spectra, when replacing $E$ and $B$ by the moment expansion of the real or imaginary parts of $\mathcal{S}_\nu$, one obtains
\begin{align}
    \mathcal{D}_\ell^{XX}(\nu)    = 
    \left(\hatpol_\nu(\bar{\beta},\bar{T})\right)^2
    &\Bigg[\mathcal{D}_\ell^{\spinscalmom_{0,X}\spinscalmom_{0,X}}
    \, + 2\mathcal{D}_\ell^{\spinscalmom_{0,X}\spinscalmom^{\beta}_{1,X}} \lnnu \nonumber\\
    &+ \mathcal{D}_\ell^{\spinscalmom^{\beta}_{1,X}\spinscalmom^{\beta}_{1,X}} \lnnu^2 \dots\Bigg],
\label{eq:DlXX}
\end{align}
with $XX\in \{EE,BB\}$. Hence, by knowing the spin-moment maps $\mathcal{W}_{k}^{\beta}$, one can in principle derive their $E$ and $B$ spectra to obtain the $\mathcal{D}_\ell^{\spinscalmom^{\beta}_{k,X}\spinscalmom^{\beta}_{m,X}}$ appearing in Eq.~\ref{eq:DlXX}. Just as with $Q$ and $U$, the $E$ and $B$ expansions are not expected to be independent, as they are the expressions of the real and imaginary parts of the same complex number $\mathcal{S}_\nu$.\footnote{To keep this link explicit, one could consider $\mathcal{D}_\ell^{\mathcal{S}\mathcal{S}}= \DLEE + \DLBB$.}  While the analysis of \citet{PlanckDust} found no significant difference between the $EE(\nu)$ and $BB(\nu)$ SEDs, a recent analysis by \citet{Ritacco22} did detect a significant difference between these SEDs in the \planck{} data. As discussed, this detection would be a direct indication of the existence of polarized mixing, which would lead to a spectral phase rotation $\EBangle(\nu)$, that is, $(\spinscalmom_{k}^{\beta})^E \neq (\spinscalmom_{k}^{\beta})^B$, and one would then expect to find a frequency dependence of the $EE/BB$ ratio in sky observations. 
The $EE/BB$ ratio can be expressed at first order as
\begin{align}
    &(r^{E/B}_\nu)_\ell \mathbf{\equiv} \frac{\DLEE(\nu)}{\DLBB(\nu)}\nonumber\\
    &=\frac{
    \mathcal{D}_\ell^{\spinscalmom_{0,E}\spinscalmom_{0,E}}
    \, + 2\mathcal{D}_\ell^{\spinscalmom_{0,E}\spinscalmom^{\beta}_{1,E}} \lnnu 
    + \mathcal{D}_\ell^{\spinscalmom^{\beta}_{1,E}\spinscalmom^{\beta}_{1,E}} \lnnu^2  }{\mathcal{D}_\ell^{\spinscalmom_{0,B}\spinscalmom_{0,B}}
    \, + 2\mathcal{D}_\ell^{\spinscalmom_{0,B}\spinscalmom^{\beta}_{1,B}} \lnnu + \mathcal{D}_\ell^{\spinscalmom^{\beta}_{1,B}\spinscalmom^{\beta}_{1,B}} \lnnu^2  }.
\label{eq:rEB}
\end{align}
We note that this expression does not depend on the modified blackbody and gives a pure ratio of moments. Therefore, looking for the $EE/BB$ ratio will probe the existence of differences between the SEDs of $E$ and $B$ due to polarized mixing independent of any choice of canonical SED for modeling dust at the voxel level. Variations of spectral parameters alone would identically distort  $E$ and $B$, leading to the same moments and leaving the $EE/BB$ ratio constant. Therefore, $(r^{E/B}_\nu)_\ell$ provides, at the power-spectra level, an observable equivalent to $\tan(\psinu)$ or $\tan(\EBangle(\nu))$ at the map level.

As in \citet{Mangilli} and \citet{Vacher2022a}, one can consider Eq.~\eqref{eq:DlXX} as two independent moment expansions for $EE$ and $BB$ and interpret the order one term as a leading-order correction to the spectral index. A scale-dependent pivot can be obtained by canceling the first order term and inserting the replacement
\begin{equation}
      \bar{\beta} \to \bar{\beta}^{XX'}_\ell = \bar{\beta} + \mathcal{D}_\ell^{\spinscalmom_{0,X}\spinscalmom_{1,X}^\beta}/\mathcal{D}_\ell^{\spinscalmom_{0,X}\spinscalmom_{0,X}}.
\end{equation}
Hence, in the presence of polarized mixing, the moments are different in $EE$ and $BB$, and it is impossible to find a common pivot simultaneously canceling the first order for $EE$ and $BB$ (i.e., $\bar{\beta}_\ell^{EE} \neq \bar{\beta}_\ell^{BB}$). 
In the perturbative regime, $\mathcal{D}_\ell^{\spinscalmom_{0,X}\spinscalmom_{0,X}} \gg \mathcal{D}_\ell^{\spinscalmom_{1,X}^\beta\spinscalmom_{1,X}^\beta} $, one can approximate Eq.~\eqref{eq:rEB} as
\begin{align}
    (r^{E/B}_\nu)_\ell 
    &\simeq A^{E/B}_\ell\fracnu^{2\Delta \bar{\beta}^{E/B}_\ell}\left(1 + \delta^{11}_\ell\lnnu^2 \right).
\label{eq:EEBBperturbative}
\end{align}
As such, while the amplitude of the power-law term $A^{E/B}_\ell= \mathcal{D}_\ell^{\spinscalmom_{0,E}\spinscalmom_{0,E}}/\mathcal{D}_\ell^{\spinscalmom_{0,B}\spinscalmom_{0,B}}$ indicates the value of the $EE/BB$ ratio at $\nu=\nu_0$, its exponent $2\Delta\bar{\beta}_\ell^{E/B}=2\bar{\beta}_\ell^{EE}-2\bar{\beta}_\ell^{BB}$ provides an indication of how the pivot spectral indices of $EE$ and $BB$ are expected to differ. The moment term, that is, 
\begin{equation}
 \delta^{11}_\ell=\frac{\mathcal{D}_\ell^{\spinscalmom_{1,E}^\beta\spinscalmom_{1,E}^\beta}}{\mathcal{D}_\ell^{\spinscalmom_{0,E}\spinscalmom_{0,E}}}-\frac{\mathcal{D}_\ell^{\spinscalmom_{1,B}^\beta\spinscalmom_{1,B}^\beta}}{\mathcal{D}_\ell^{\spinscalmom_{0,B}\spinscalmom_{0,B}}},
\end{equation}
quantifies the difference between the auto-correlation of the order one spin-moments in $EE$ and $BB$ and cannot be strictly equal to zero if the $EE/BB$ ratio is spectral dependent.

\subsubsection{The \texorpdfstring{$EB$}{EB} power spectrum \label{sec:EB}}

A similar calculation can be done for the cross $EB$ spectra, leading to
\begin{align}
    \DLEB(\nu)    = \left(\hatpol_\nu (\bar{\beta},\bar{T})\right)^2 \Bigg[&\mathcal{D}_\ell^{\spinscalmom_{0,E}\spinscalmom_{0,B}}
    \,\nonumber \\
    &+ \left(\mathcal{D}_\ell^{\spinscalmom_{0,E}\spinscalmom^{\beta}_{1,B}} + \mathcal{D}_\ell^{\spinscalmom_{0,B}\spinscalmom^{\beta}_{1,E}}\right) \lnnu \nonumber\\
    &+ \mathcal{D}_\ell^{\spinscalmom^{\beta}_{1,E}\spinscalmom^{\beta}_{1,B}} \lnnu^2 \dots\Bigg].
    \label{eq:eb_moments}
\end{align}
The zeroth order term solely quantifies the structure of the magnetized interstellar medium, {as it depends only on the maps of $\mathcal{W}_{0}$ and hence on the distribution of the matter density and magnetic field orientations}. From parity considerations, this term is expected to be very small \citep{PlanckDust2,Clark2021}. Even if the leading term in Eq~\eqref{eq:eb_moments} is null, a nonzero frequency dependent $\DLEB$ can be generated by the two other terms, which follow from variations of dust emission properties. 

Isolating the $EB$ moments is not a trivial task, as, for example, the $EB/EE$ quantity could be dominated by the $EE$ distortions. 
The favorable option would be to analytically correct for the scale dependent pivot as
\begin{equation}
      \bar{\beta} \to \bar{\beta}^{EB}_\ell = \bar{\beta} + \left(\mathcal{D}_\ell^{\spinscalmom_{0,E}\spinscalmom_{1,B}^\beta}+\mathcal{D}_\ell^{\spinscalmom_{0,B}\spinscalmom_{1,E}^\beta} \right)/(2\mathcal{D}_\ell^{\spinscalmom_{0,X}\spinscalmom_{0,X}}),
\label{eq:pivotEB}
\end{equation}
assuming that the first term is the dominant order, which should always be true if the mean signal is in the perturbative regime. Here again, in the presence of polarized mixing, we observe that $\bar{\beta}^{EB}\neq \bar{\beta}^{EE} \neq \bar{\beta}^{BB}$ such that the three polarized spectra will have a different effective SED. This highlights that observing a spectral dependence of the $EE/BB$ ratio guarantees the existence of $EB$ distortions at some level.

However, in observational conditions, we cannot analytically compute  the pivot defined in Eq.~\eqref{eq:pivotEB}, as we do not have access to the 3D distribution of spectral parameters and polarization angles. Therefore, in order to highlight the $EB$ SED distortions, one can choose any $\bar{\beta}^{EB}$ and consider the ratio
\begin{equation}
    (r_\nu^{E\times B})_\ell \mathbf{\equiv} \frac{\DLEB}{\left(\hatpol_\nu (\bar{\beta}_\ell^{EB},\bar{T})\right)^2}.
\label{eq:rExB}
\end{equation}
The amplitude of the variations of $(r_\nu^{E\times B})_\ell$ will depend strongly on the choice of $\bar{\beta}_\ell^{EB}$. Thus, one could imagine  fitting $\bar{\beta}^{EB}_\ell$ directly onto the $EB$ data and minimizing the amplitude of the distortions.  But as the $EB$ Galactic signal is very low, this fit might dramatically increase the Galactic modeling uncertainty. One could also use a proxy for $\bar{\beta}_\ell^{EB}$ (for example, from the high signal-to-noise $\bar{\beta}_\ell^{EE}$), but this will result in enhancing the $EB$ distortions if $|\bar{\beta}_\ell^{EB}-\bar{\beta}_\ell^{EE}|\gg0$. Still, since $EE$, $BB$, and $EB$ have a common physical origin, they should be treated together with a shared $\bar{\beta}$ in the spin-moment formalism.

\section{The toy model filament}
\label{sec:toy-model}

In order to refine the toy model of the infinite filament presented in Sec.~\ref{sec:EBmix}, 
we consider again an infinite filament in front of a polarized background  having both an MBB emission law with different spectral indices and polarization angles. The frequency dependence of the polarization angle $\psi$ field in the filament would arise naturally from the polarized mixing described in Sec.~\ref{sec:QUmix} (see Fig.~\ref{fig:toy-model}). We chose $\psibg=0^{\circ}$ and considered various cases for $\psifl$ (where the {superscripts} bg and fl stand for background and filament, respectively). 
From astrophysical considerations, one {might} expect filaments to be colder than the diffuse background, that is, $T^{\rm fl} < T^{\rm bg}$. 
Here again, for the sake of simplicity, we do not consider temperature variations, fixing $\beta^{\rm fl}=1.8$, $\beta^{\rm bg}=1.5$, and $\bar{T}=T^{\rm fl} = T^{\rm bg}= 20$\,K. We also used  $A^{\rm fl}=A^{\rm bg} = 1$, assuming that the background and the filament share the same optical depth and the same inclination with respect to the Galactic magnetic field. 
In order to keep the analysis easy to interpret, we also ignored the impact of the size and orientation of the filament. Changing these parameters is expected to change the relative amplitudes of the spectra \citep{Huffenberger2020}, but our conclusions should not be impacted regarding the moment expansion formalism and the impact of polarized mixing on the angular power spectra.

The toy model map is a $32\times32$ flat pixel grid on which the filament represents a $11\times 32$ vertical rectangle (see Fig.~\ref{fig:toy-model}). The filament is still assumed to be infinite, as the power spectra computation assumes periodic boundary conditions. We treated this example numerically using the {\sc Namaster} library \citep{namaster} in order to evaluate the polarized power spectra $\mathcal{D}^{XX'}_{\ell}$ of the flat sky maps in a single multipole bin containing the 15 first values of $\ell$. The frequency range was chosen to be an array from 100 to 400\,GHz with intervals of 10\,GHz and spanning a frequency interval relevant for CMB missions, under which the effect of the spectral index variations is expected to be dominant over possible temperature variations \citep{Ptep}. The reference frequency was chosen to be $\nu_0=400$\,GHz. 

\subsection{Single pixel analysis}
\begin{figure*}[t!]
    \centering
    \includegraphics[scale=0.55]{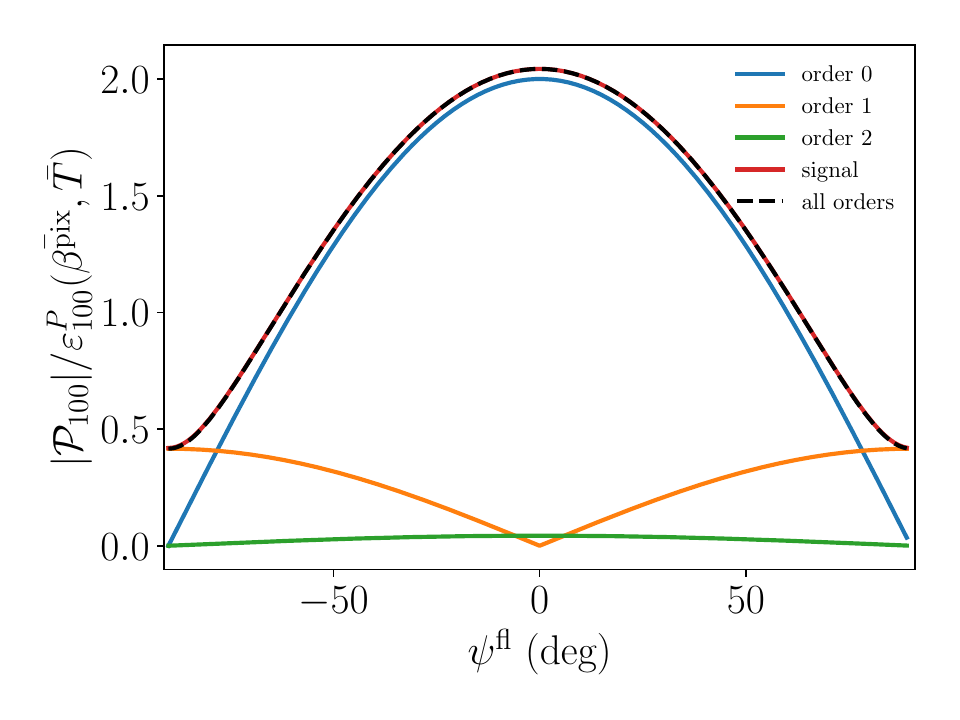}
    \includegraphics[scale=0.55]{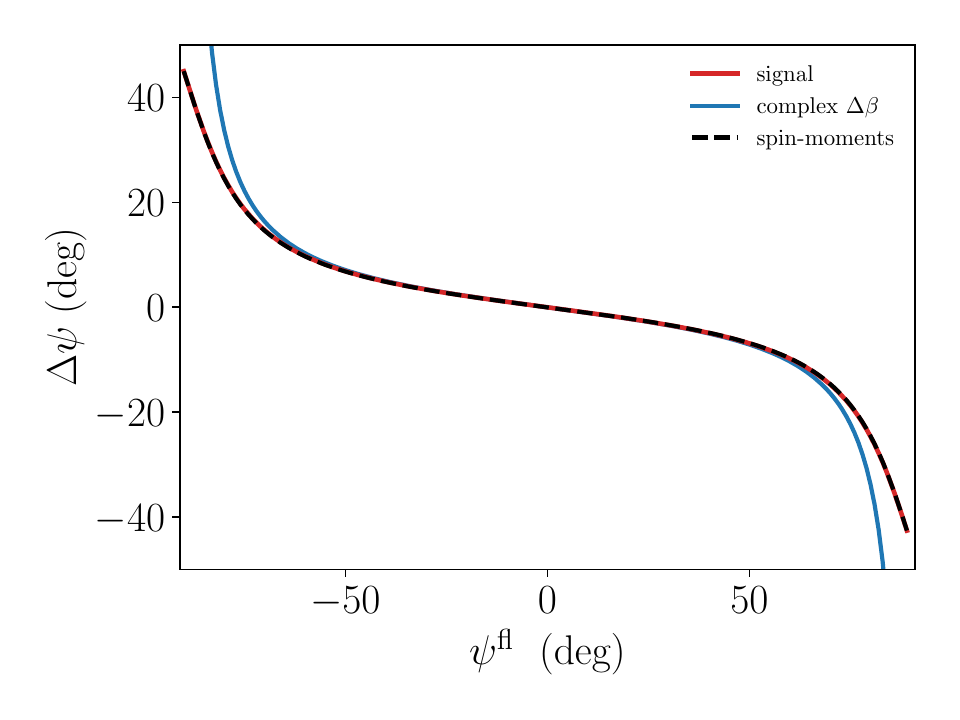}
    \caption{Total polarization spinor in a single pixel of the filament of the toy model for $\psibg=0^\circ$ and various values of the filament polarization angle $\psifl$. Left: Modulus of the total signal at $100$\,GHz normalized by the pivot MBB (red) and modulus of the analytical derivation from the spin-moment expansion up to second order (black dashed line). The modulus of each term is displayed: order 0: $|\mathcal{W}_0|$ (blue), order 1: $|\mathcal{W}^\beta_1\ln\left(100/400\right)|$ (orange), and  order 2: $|0.5\mathcal{W}^\beta_2\ln\left(100/400\right)^2|$ (green). Right: Difference of the polarization angles between the two frequencies. Signal (red), prediction from the complex $\Delta\beta$ correction $0.5\,{\rm Im}(\mathcal{W}^\beta_1/\mathcal{W}_0)\ln\left(100/400\right)$ (blue) and from the spin-moment expansion up to second order (black dashed line).}
    \label{fig:signalfilament}
\end{figure*}

\begin{figure*}[t!]
    \centering
    \includegraphics[scale=0.55]{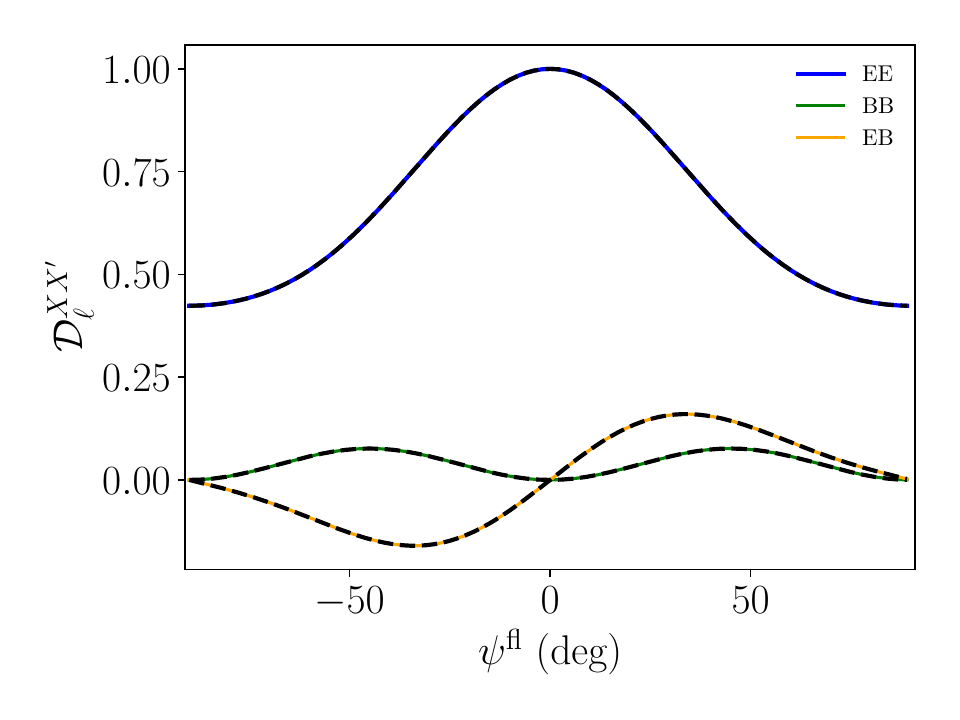}
    \includegraphics[scale=0.55]{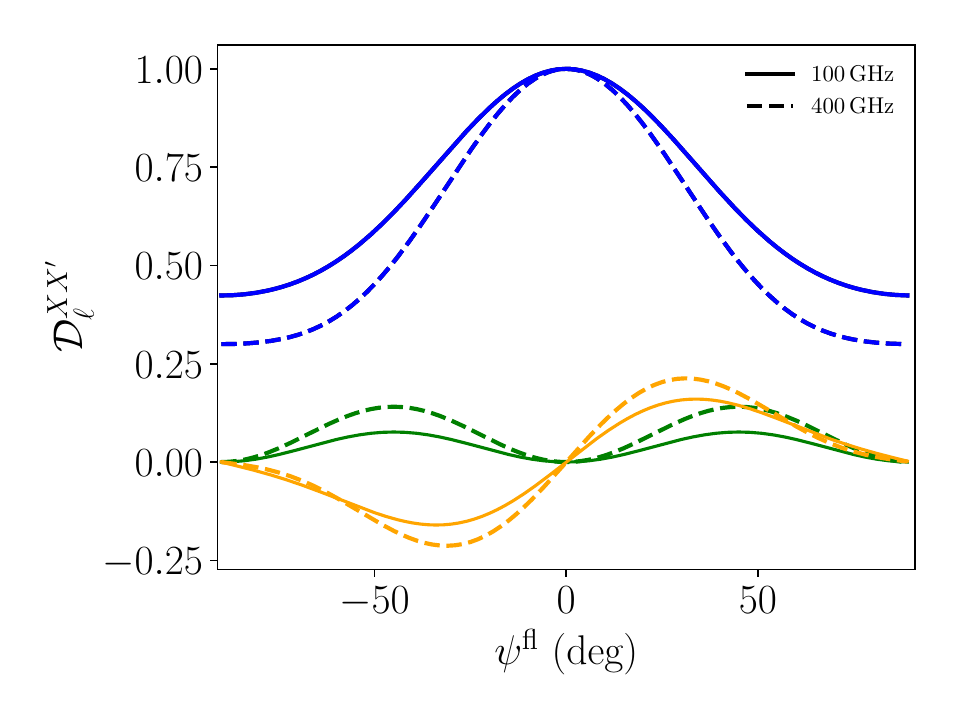}
    \caption{Mean polarized power spectra for the toy model filament with $\psibg=0^\circ$ and various values of the filament polarization angle $\psifl$. The $EE$ (blue), $BB$ (green), and $EB$ (orange) angular power spectra are given at two different frequencies: 100 (continuous line) and 400 GHz (dashed line). Each spectra was normalized by the maximum value of $\left(\left(\DLEE\right)^2(\nu) + \left(\DLBB\right)^2(\nu)\right)^{1/2}$. Left: No SED distortions: $\beta^{\rm fl}=\beta^{\rm bg}=1.5$. Right: with SED distortions: $\beta^{\rm fl}=1.8$ and $\beta^{\rm bg}=1.5$.}
    \label{fig:EB-filament}
\end{figure*}  


In this section, we attempt to evaluate how the geometrical and spectral aspects of the signal are intertwined to produce the resulting angular power spectra. As a first step, we focus on the benefits of the spin-moment approach in a single filament pixel. 

We first considered only the two frequencies $\nu_1 = 100$\,GHz and $\nu_2=400$\,GHz and tested how the results changed under a variation of $\psifl$ in the range $[-90^{\circ},90^{\circ}]$.
The modulus $|\mathcal{P}_\nu|$ and phase $\psinu$ of the total polarization spinor in the filament are displayed in Fig.~\ref{fig:signalfilament}. In the left panel, we display the departures from the pivot modified blackbody of the total signal modulus in the filament at $100$\,GHz, $|\mathcal{P}_{100}|/|\hatpol_{100}(\bar{\beta}^{\rm pix},\bar{T})|$ with $\bar{\beta}^{\rm pix}= (\beta^{\rm fl}+ \beta^{\rm bg})/2=1.65$. These departures were well modeled by the spin-moment expansion up to order two, which one can derive analytically using Equations \eqref{eq:spin-moment-expansion} and \eqref{eq:spin-moments}, for every value of $\psifl$. 
The modulus of the signal appears to become smaller when $\psifl$ goes away from $0^{\circ}$. This is due to a corresponding vanishing of the amplitude of $\mathcal{W}_0= A_{\rm fl}e^{2\i\psifl} + A_{\rm bg}e^{2\i\psibg}$, which is due to progressive depolarization. Indeed at $\psifl=0^{\circ}$, the phases are aligned, and the complex amplitude reaches its maximal value: $\mathcal{W}_0= A_{\rm bg} + A_{\rm fl} = 2$. Moving away from $\psifl=0^{\circ}$, the phases cancel each other down to $\mathcal{W}_0=A_{\rm bg}e^{2\i 0} + A_{\rm fl}e^{\pm2\i \pi/2 } = 0$. We note that this is a pure geometrical "spin-2" effect independent of the values of the spectral parameters. In contrast, the first order moment increases when going away from $\psifl=0^{\circ}$, producing an increase of the distortion amplitudes and a change of polarization angle with frequency $\Delta \psi = \psi(400)-\psi(100)$, as shown in the right panel of Fig.~\ref{fig:signalfilament}. The behavior of $\Delta \psi$ with $\psifl$ is well modeled by the complex pivot correction  (Eq.~\eqref{eq:psi_nu_lnnu} and \eqref{eq:psi_nu_lnnu2}), except in the nonperturbative regime when $\mathcal{W}^{\beta}_1$ becomes large compared to $ \mathcal{W}_0$, as discussed in \citet{Vacher2022b}. In this regime, a pivot correction is impossible, and all the terms of the expansion must be kept. We note that the second moment $\mathcal{W}_2^\beta$ is still required to obtain a good model for $|\mathcal{P}_\nu|$ around $\psifl=0^\circ$ when it is greater than the first order ($\mathcal{W}^\beta_2 > \mathcal{W}^\beta_1 \simeq 0$). Indeed, as already discussed in Sec.~\ref{sec:QUmix}, it is impossible to always guarantee the hierarchy between the moments due to the complex nature of the expansion and its nonperturbative treatment of polarization angles. Overall, both the modulus and the phase of the complex signal in the pixel are fully predicted by the spin-moment expansion. 
\begin{figure*}[t]
    \centering
    \includegraphics[scale=0.55]{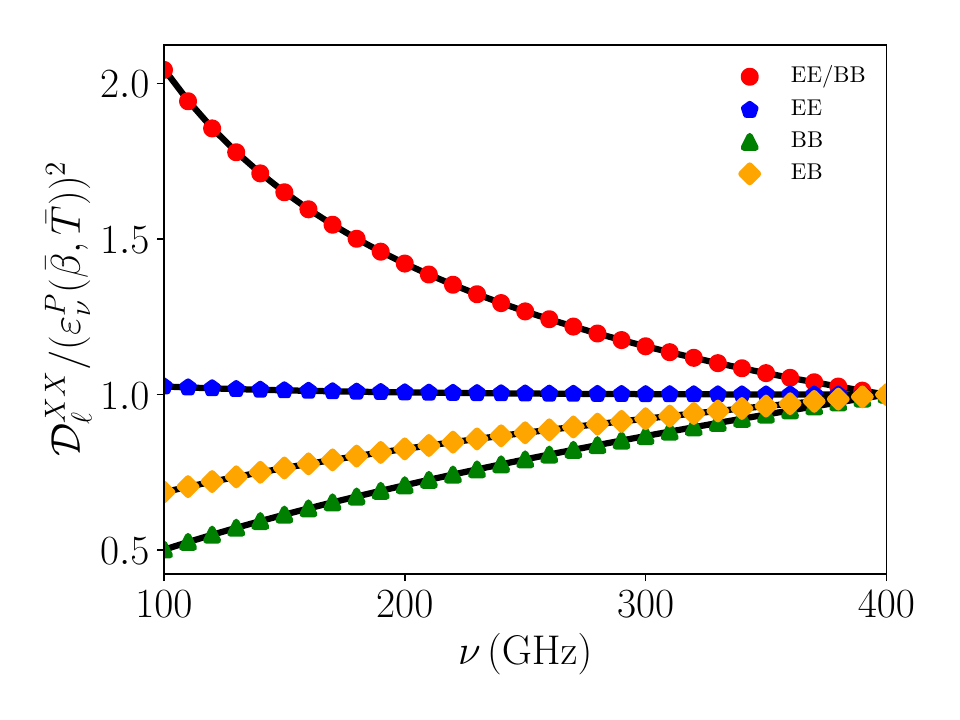}
    \includegraphics[scale=0.55]{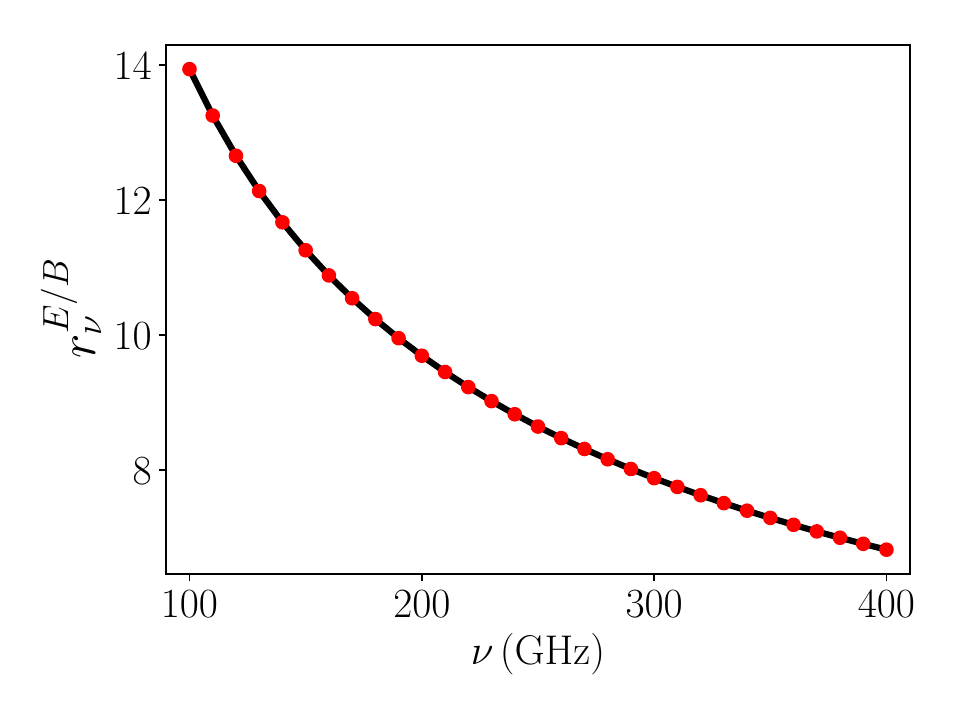}
    \caption{Spectral dependence of the toy model filament polarized power spectra. Left: Polarized power spectra divided by the pivot-modified blackbody squared for the special case of $\psi^{\rm bg}=30^{\circ}$. All spectra are normalized with respect to their value at $\nu_0$. The signals are shown in color: $EE/BB$ (red), $EE$ (blue), $BB$ (green), and $ EB$ (orange). The black dashed lines were recovered by deriving the moment power spectra from the spin-moment maps. Right: Close up view on $r_\nu^{E/B}$. The black curve represents the best-fit of Eq.\eqref{eq:rEB}.}
    \label{fig:PS-freq}
\end{figure*}


\subsection{Power spectra analysis}
As a second step, we studied the behavior of the power spectra, displayed in Fig.~\ref{fig:EB-filament}, for the same filament model. In order to compare the frequencies, which have orders of magnitudes of differences, we normalized all the spectra by the maximal value of $\left(\left(\DLEE\right)^2(\nu) + \left(\DLBB\right)^2(\nu)\right)^{1/2}$ at each frequency. In the case without distortions, when $\beta^{\rm fl}= \beta^{\rm bg}=1.5$, all the polarized angular power spectra displayed an identical behavior between the two frequencies, as the $E$- and $B$-modes shared the same SED. Hence, no moments were expected in either  $\mathcal{P}_\nu$ or in $\mathcal{S}_\nu$. The overall behavior of the angular power spectra with respect to the filament's angle displayed in Fig.~\ref{fig:EB-filament} is very similar to the magnetic misalignment phenomenon (see Fig.~2 of \citet{Clark2021}), considering the extra depolarization effect. At $\psifl=0^{\circ}$, the sum of the MBBs in the filament is aligned with the MBB in the background. The $E$-modes are hence maximal, and there is no $B$-mode or $EB$ correlation. For $\psifl=\pm 45^{\circ}$, $B$-modes are maximal but lower than the maximum of the $E$-modes due to the progressive depolarization and corresponding amplitude loss discussed above. At $\psifl=\pm 90^{\circ}$, the $E$-modes are expected to peak again, but they did not due to depolarization that makes the signal minimal, and $B$-modes returned to zero. The absolute value of $EB$ is maximal when $\sqrt{EE^2+BB^2}$ is maximal.

When considering our example with $\beta^{\rm fl} \neq \beta^{\rm bg}$, distortions appeared, as indicated by the rotation of $\psi$ between 100 and 400\,GHz in Fig.~\ref{fig:signalfilament}. 
No matter what the value of $\psifl$ was, $\psi$ drifted away from alignment between the two frequencies (according to Fig. \ref{fig:signalfilament}, with a positive angle for $\psifl>0$ and negative for $\psifl<0$), leading to an increase of the $E$-modes and a corresponding decrease of the $B$-modes from $100$ to $400$ GHz such that one would expect $r^{E/B}_\nu$ to decrease with frequency.
The distortion was greater as $\psifl$ went away from 0$^\circ$, which is in agreement with the values of $\mathcal{W}_1^\beta$ in Fig.~\ref{fig:signalfilament}. 
Distortions were also witnessed for the $EB$ spectra, illustrating how the mixing of polarized signals can increase the amplitude of this parity, thus violating spectra from one frequency to another.
Changing the value of $\psibg$ would not change the above conclusions but would change the relative amplitudes of the $EE$, $BB$, and $EB$ angular power spectra. In order to remain concise, other cases with $\psibg=45^{\circ}$ ($B$-modes maximum) and $\psibg=22.5^{\circ}$ (equipartition of $E$- and $B$-modes) are displayed in Appendix~\ref{sec:Appendix-psifg}. 
\begin{figure}[h]
    \centering
    \includegraphics[width=\linewidth]{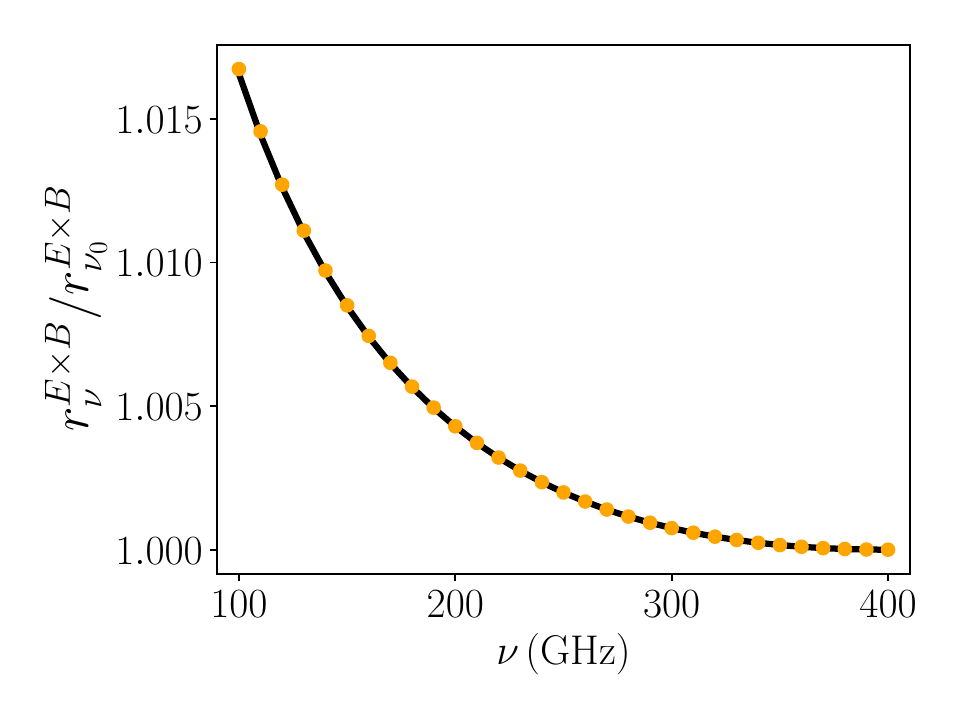}
    \caption{ Graph of $r_\nu^{E\times B}$ normalized at $\nu_0$ for the toy model filament. The black curve represents the analytical prediction from the spin-moment maps.}
    \label{fig:toy-model-EB}
\end{figure}
\begin{table*}
    \centering
    \caption{Best-fit and predicted values for the parameters entering in the perturbative expression of $\left( r_{\nu}^{E/B}\right)_{\ell}.$ (See Eq.~\eqref{eq:EEBBperturbative} for the case of the toy model filament.)}
    \begin{tabular}{l|c|c|c}
\toprule
             &$ A^{E/B}_\ell$ & $ 2\Delta \bar{\beta}^{E/B}_\ell $ & $\delta^{11}_\ell$  \\
\toprule
                {{Prediction } }& $6.818$ & $-0.4977$ & $0.00178$  \\
\midrule
                {{Best fit } }& $6.818  \pm  0.0001  $ & $  -0.4986 \pm 0.0001$ & $0.01251 \pm 0.00006$\\
\bottomrule
\end{tabular}
\label{tab:EEBB}
\end{table*}

\subsection{Predicting the spectral dependence of the power spectra. \label{sec:cl-spectral-dep}}

The spectral behavior of the angular power spectra discussed above can be predicted by the spin-moment expansion.
From Eq.~\eqref{eq:spin-moments}, one can build maps of the spin-moments by knowing the $\psi$ and $\beta$ distributions in each pixel and using, as a common pivot, the mean spectral parameter over the whole map $\bar{\beta} = \langle A\beta\rangle/\langle A \rangle \sim 1.55$. It is then straightforward to compute the corresponding $\mathcal{D}_\ell^{\spinscalmom_{k,X}\spinscalmom_{k,X'}}$ for each pair of moments. To do this, we again used {\sc Namaster} on the flat sky moment maps. 
We then obtained the frequency dependent power spectra by inserting the $\mathcal{D}_\ell^{\spinscalmom_{k,X}\spinscalmom_{k,X'}}$ in Eqs.~ \eqref{eq:DlXX} and \eqref{eq:eb_moments}. 
Figure~\ref{fig:PS-freq} shows a comparison between the spin-moment prediction up to order three and the signal over the whole frequency range for the special case $\psi^{\rm bg}=30^{\circ}$. These examples demonstrate that the expansion we derived is correct, and that it is possible to infer the polarized power spectra from the spin-moment maps, which themselves are derived from the spectral parameter and polarization angle distributions. In experimental conditions, however, one cannot directly access the distributions of spectral parameters and polarization angles, making this  derivation impossible. We show, however, that the spin-moments and their expansion at the power-spectra level provide robust models for an accurate characterization of the polarized signal regardless of the distributions of $\beta$ and $\psi$. A detailed study of how far one can go  by inferring the dust properties from the power spectra and/or spin-moment maps is left for future work. {First steps toward that direction can be found in recent works such as \citet{Sponseller2022} and \citet{McBride2022}, which are aimed at quantifying the resulting biases obtained on the recovered CMB and dust parameters, depending on the underlying probability distributions of the spectral parameters integrated into the signal.}

In order to assess the validity of the $EE/BB$ ratio approximation in Eq.~\eqref{eq:EEBBperturbative}, we fit the $EE/BB$ ratio with a weighting proportional to the signal itself using the {\sc lmfit} software \citep{Lmfit}. In Fig.~\ref{fig:PS-freq}, good agreement between the fit and the data points can be observed. The best-fit values of $A_\ell^{E/B}$, $2\Delta \bar{\beta}^{E/B}_\ell$ and $\delta^{11}_\ell$ can be found in Tab.~\ref{tab:EEBB}. We compared the values  to the ones predicted using the moment maps and from computing the pivots $\bar{\beta}_\ell^{EE}$. The values are close but not equal because Eq.~\eqref{eq:EEBBperturbative} stops at order one and the fit compensates for the higher-order moments. Still, the expression provides a simple and interpretable model to characterize the spectral dependence of the $EE/BB$ ratio.

Finally, as discussed in Sec.~\ref{sec:EB}, we used a pivot $\bar\beta^{EB}$ in order to evaluate $r_\nu^{E\times B}$. Both fitting and analytical derivations allowed us to find $\bar{\beta}^{EB}_\ell\sim 1.69 \neq \bar{\beta}$. The ratio $r_\nu^{E\times B}$ is displayed in Fig.~\ref{fig:toy-model-EB}, quantifying the amplitude of the residual distortions of higher orders. The variations are at the percent level. Once again, the prediction from moment expansion overlaps with the signal, validating our methodology. 
\begin{figure*}[t]
    \centering
    \includegraphics[scale=0.35]{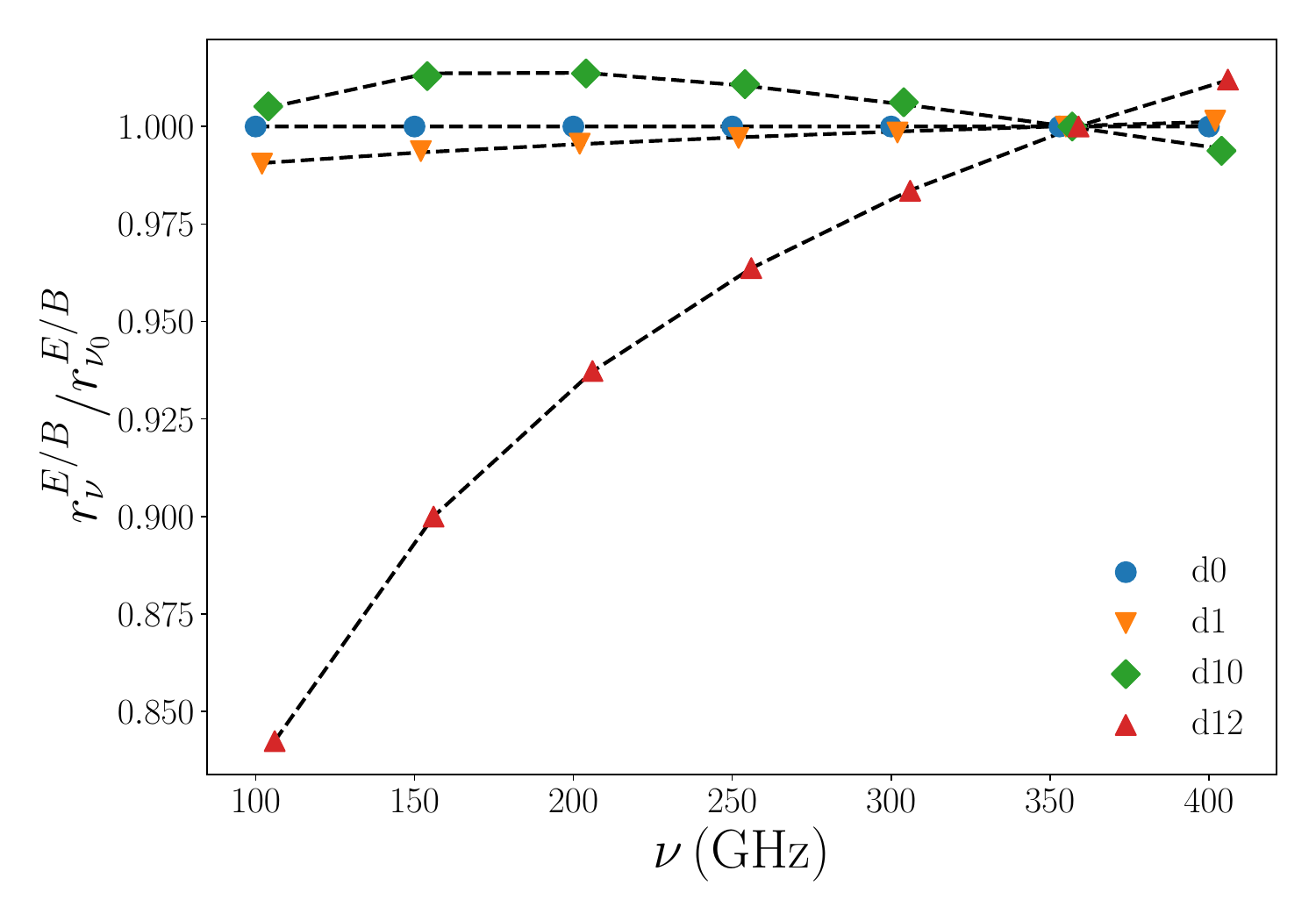}
    \includegraphics[scale=0.35]{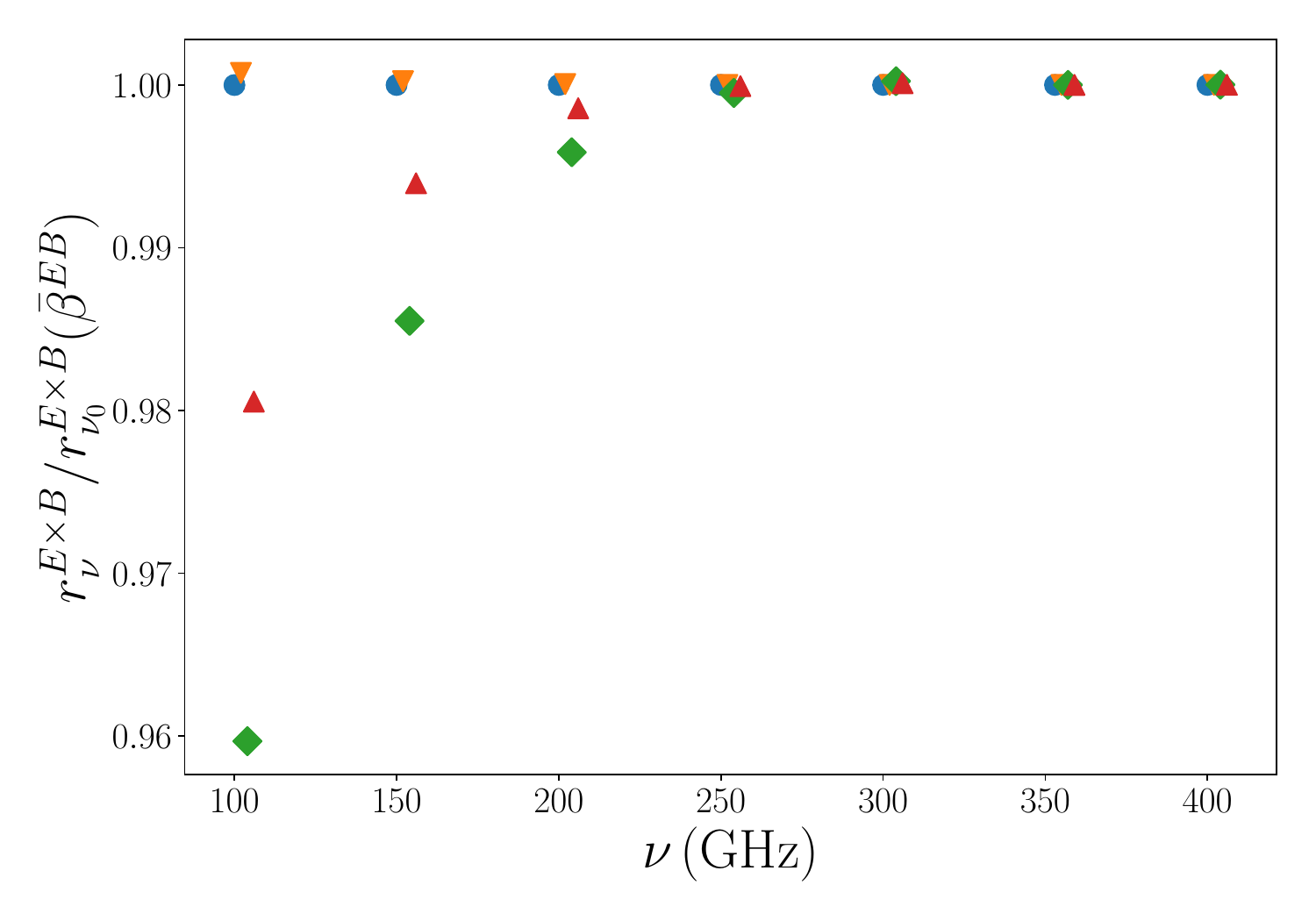}
    \caption{Graph of $r^{E/B}_\nu$ and $r^{E\times B}_\nu$ for different {\sc PySM} dust models. The shapes are defined as follows:\ \dzero{} (blue circles), \done{} (orange reversed triangles), \dten{} (green diamonds), and \dtwelve{} (red triangles). }
    \label{fig:alldust_EEBB_EBfitEB}
\end{figure*}

\section{{\sc PySM} 3 models \label{sec:Pysm}}
In this section, we illustrate the previously discussed phenomenon on the {\sc PySM} 3 models\footnote{{The version {\tt 3.4.0b4} of the {\sc PySM} 3 was used in this work.}} \citep{Pysm,Zonca_2021}. Computations are made on the sphere using the {\sc healpy} package \citep{healpix}. 
We considered the four following models:
\begin{itemize}
    \item \dzero{}: This model was built with polarization $Q$ and $U$ map templates from  the \planck{} 2015 data at 353\,GHz \citep{planck_2015_overview} and extrapolated at all frequencies using a single MBB with a fixed $\beta=1.54$ and $T=20$\,K over the sky.
    \item \done{}: This model uses the same $Q$ and $U$ map templates as \dzero{} and was extrapolated to other frequencies using an MBB with a varying $\beta$ and $T$ between pixels across the sky.
    \item \dten{}: This model is a refined version of \done{}. The extrapolation was performed using an MBB with spectral parameters in each pixel coming from templates of the Generalized Needlet Internal Linear Combination (GNILC) needlet-based analysis of \planck{} data \citep{GNILC2011} and includes a color correction and random fluctuations of $\beta$ and $T$ on small scales.
    \item \dtwelve{}: This model was built out of six overlapping MBBs, as detailed in \citet{Martinez-Solaeche2018}. This is the only model that has variations of the spectral parameters and polarization angles along the line of sight, which is the same as in the toy model filament presented in Sec.~\ref{sec:toy-model}.
\end{itemize}

We again chose the frequency interval $\nu\in\{100,400\}$\,GHz, with  steps of 50\,GHz replacing the value at $350$ GHz by the reference frequency of the models $\nu_0 =353$\,GHz. Power spectra were computed again using {\sc Namaster} in a single multipole bin from $\ell_0=2$ to $\ell_{\rm max}=200$ at a {\sc healpy} resolution of $N_{\rm side}=128$.
In order to observe a large patch of the sky while still avoiding the central Galactic region, we used the \planck{} {\sc GAL080} raw mask with $f_{\rm sky}=0.8$ available on the \href{https://pla.esac.esa.int/#maps}{\planck{} Legacy Archive}. We subsequently performed a {\sc Namaster} C2 apodization with a scale of $2^{\circ}$. Both the $E$- and $B$-modes were purified during the spectra computations.

In principle, by knowing the $A$, $\psi$, $\beta$, and $T$ templates of the {\sc PySM} maps, it is possible to analytically compute the spectra expansion as we did in Sec.~\ref{sec:cl-spectral-dep}. This would however require consideration of the  temperature effects and the $\beta-T$ correlations, which are expected to have a significant impact on the modeling, as discussed in \citet{Vacher2022a} and \citet{Sponseller2022}. 

\subsection{The \texorpdfstring{$EE/BB$}{EE/BB} ratio}

In this section, we first focus on the $EE/BB$ ratio. The ratio $r_\nu^{E/B}$ is displayed in the left panel of Fig.~\ref{fig:alldust_EEBB_EBfitEB}. As expected, no departure from constancy is observed for the $EE/BB$ ratio in the case of \dzero{}. {We note that this result would remain valid even if the canonical SED was not given by an MBB, as long as the associated spectral parameters remained constant over the sky.}\footnote{{An example would be given by the $\deight$ model where SED is given by an adjusted version of the model proposed in \citet{Hensley2017} with constant spectral parameters across the sky.}}
A frequency dependent EE/BB ratio is expected only in the presence of polarized mixing, that is, between or along the lines of sight. As previously stressed, this makes the $EE/BB$ ratio a powerful probe of these variations independent of the SED effectively used to locally describe the signal.

For $\done{}$ and $\dten{}$, we found variations on the order of a few percentage points. Even if the SEDs in each pixel are nondistorted MBBs, variations between lines of sight are enough to produce a frequency dependence of the $EE/BB$ power spectra. The $\dtwelve{}$ model, which contains both variations of the spectral properties along and between the lines of sight, displayed stronger variations of up to 15\%.

A fit of Eq.~\eqref{eq:EEBBperturbative} for each model is also displayed in Fig.~\ref{fig:alldust_EEBB_EBfitEB}. The resulting curve from the fit appeared to be in good agreement with the signal in all cases, ensuring that the perturbative expression proposed in Sec.~\ref{sec:EE-BB} remains a good way to quantify the departures from constancy of $EE/BB$ on realistic dust templates.

In all cases, the observed amplitudes of variations were expected to change widely depending on the sky fraction and the range of multipoles considered, as averages were made over different Galactic regions. {As spectral parameters and magnetic field orientations are expected to have distinct scale dependencies, averaging over different multipoles when binning spectra represents an additional source of distortions.} The impact of such {variations of the $EE$/$BB$ ratio} on cosmological analysis is left for future work, but we expect it to be substantial for CMB $B$-mode analyses, as mismodeling of the dust component by a few percentage points will have a significant impact when trying to reach a measurement, for example, of $r=0.001$ \citep{PlanckDust2}.
\begin{figure*}[t]
    \centering
    \includegraphics[scale=0.35]{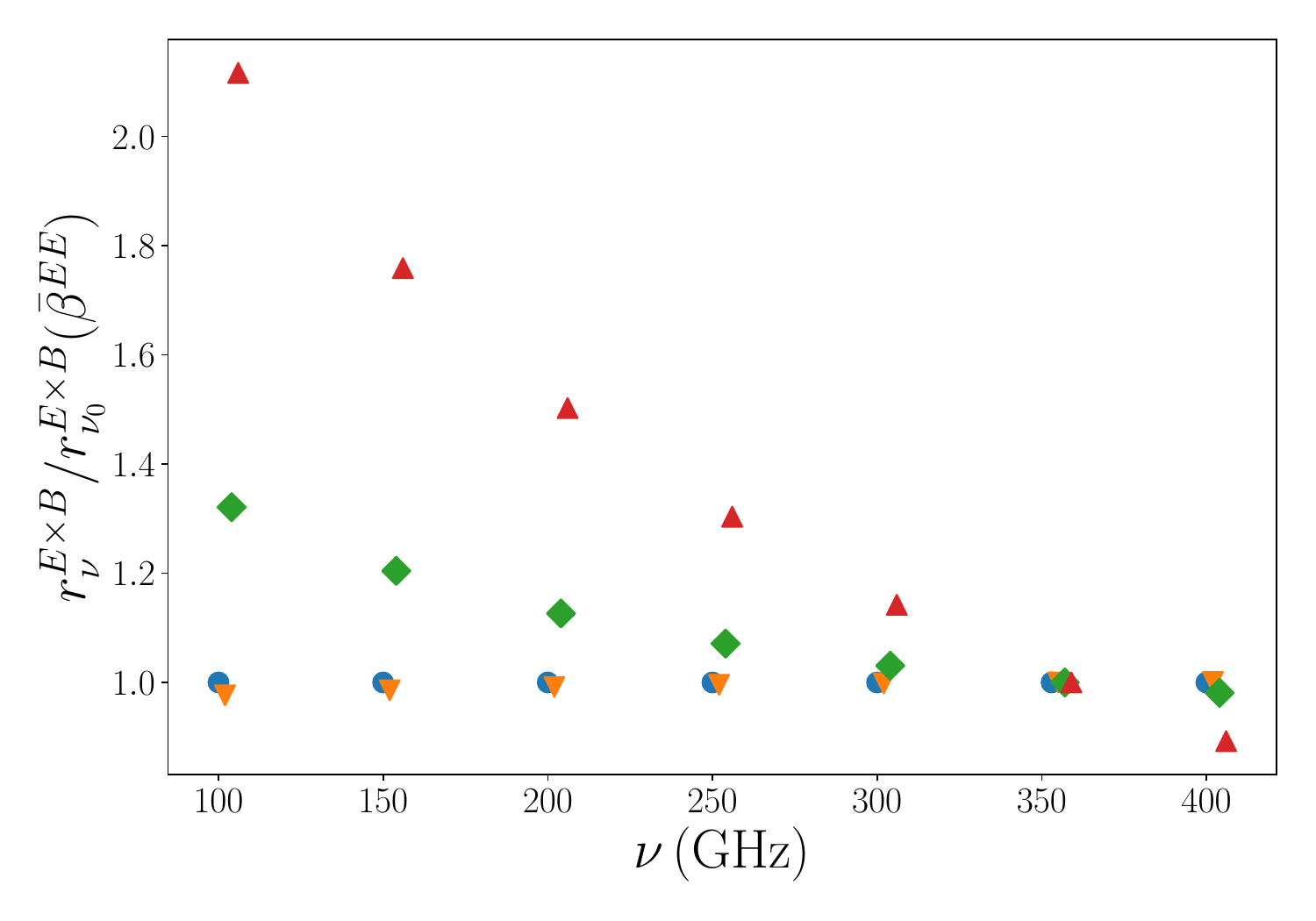}
    \includegraphics[scale=0.35]{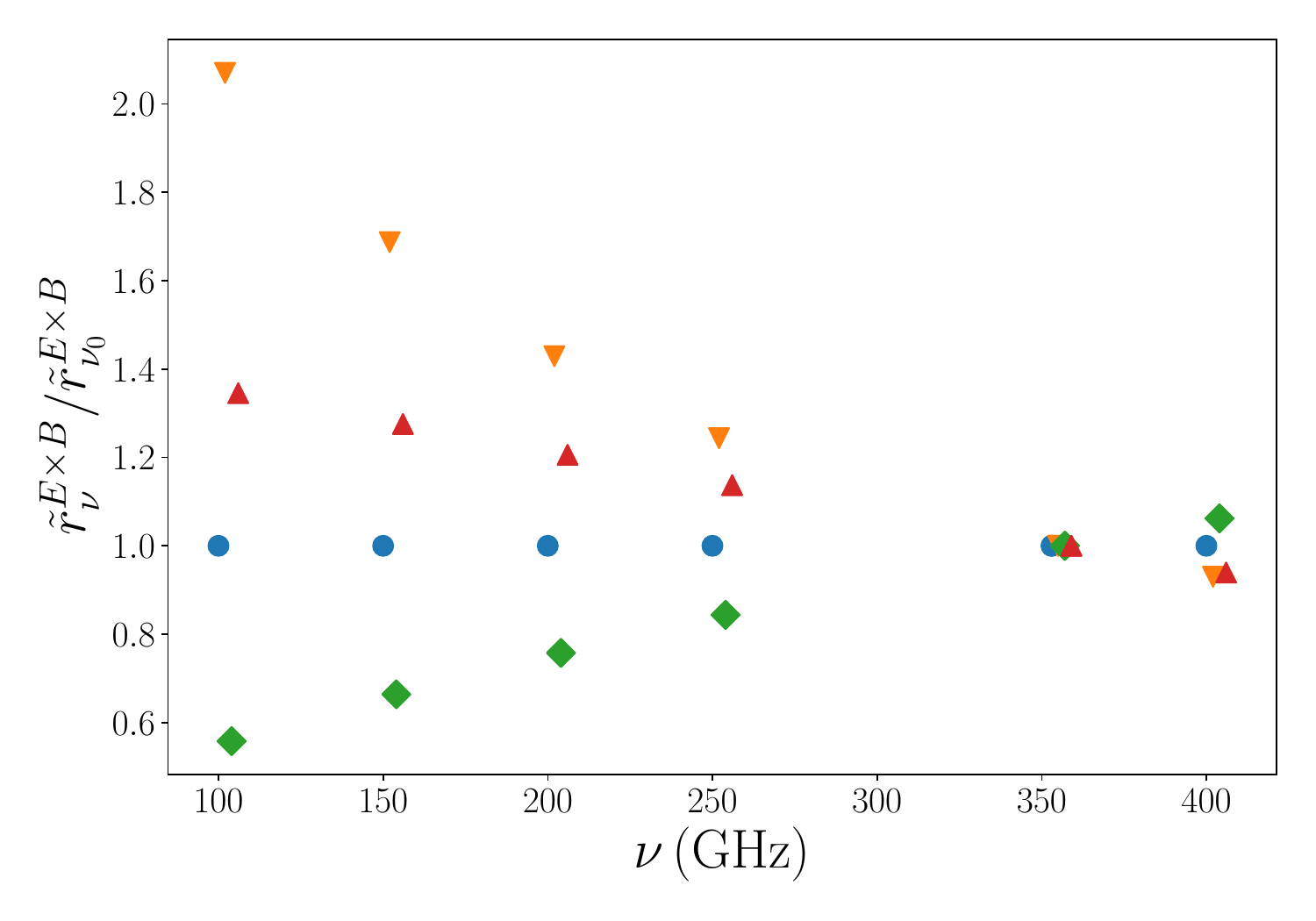}
    \caption{Graph of $r^{E\times B}_\nu$ and $\tilde{r}^{E\times B}_\nu$ 
    for different {\sc PySM} dust models. The symbols are defined as follows:\ \dzero{} (blue circles), \done{} (orange reversed triangles), \dten{} (green diamonds), and \dtwelve{} (red triangles).}
    \label{fig:alldust_cltilde}
\end{figure*}

\subsection{The \texorpdfstring{$EB$}{EB} spectra and cosmic birefringence}

As discussed in Sec.~\ref{sec:EB}, highlighting the $EB$ distortions is not trivial. In order to do so, we first performed a fit of $\bar{\beta}^{EB}$ and $\bar{T}^{EB}$ directly on the signal and considered $r_\nu^{E\times B}$. The results are displayed in the right panel of Fig.~\ref{fig:alldust_EEBB_EBfitEB}. The changes with frequency at the percent level for all but the \dzero{} model are clear, indicating the presence of distortions at a level that could be neglected for these sky models in contemporary birefringence analysis.  

As discussed in Sec.~\ref{sec:EB}, however, because the dust $EB$ signal is currently very small compared to the measurement errors in real observational conditions (and will probably remain modest in the future), it is impossible to access these quantities as directly as we did here, and one would instead be tempted to use the high signal-to-noise $\bar{\beta}^{EE}$ and $\bar{T}^{EE}$ as proxys for the $EB$ spectra. In the left panel of Fig.~\ref{fig:alldust_cltilde}, we show $r_\nu^{E \times B}$ using the best fits of the $EE$ spectral parameters as a pivot. The figure shows the existence of spectral variations from a few percentage points for the \done{} model to approximately 40\% for the \dten{} model and up to a factor of two  for the \dtwelve{} model. As such, the choice of spectral parameters used to highlight the $EB$ SED is extremely important and requires particular attention.

It is also relevant to consider the quantity 
\begin{equation}
    \widetilde{\mathcal{D}}_\ell^{EB} \simeq A^{E\times B}_\ell \frac{\mathcal{D}_\ell^{EE}\mathcal{D}_\ell^{TB}}{\mathcal{D}_\ell^{TE}},
\label{eq:Dtilde}
\end{equation}
which can provide a higher signal-to-noise estimator of the foreground $EB$ signal \citep{Clark2021,Diego-Palazuelos2022a}. As a scale dependent amplitude, $A^{E\times B}_\ell$ is frequency independent by construction. In order to quantify the deviations to this approximation, we considered the ratio 
\begin{equation}
 (\tilde{r}_\nu^{E\times B})_\ell = \frac{\mathcal{D}_\ell^{EB}}{\widetilde{\mathcal{D}}_\ell^{EB}}.
 \label{ExB_model}
\end{equation}
Results are presented in Fig.~\ref{fig:alldust_cltilde}. Large departures from Eq.~\ref{eq:Dtilde} can be observed away from $\nu_0$ for all the models but the \dzero{} model. According to the moment formalism presented above, the SEDs of $EE$, $TB$, and $TE$  present distinct distortions in the presence of polarized mixing and are in turn different from that of $EB$, thus explaining why this approximation becomes invalid when at different frequencies. For an accurate modeling of the parity violating foreground signal and in order to probe the existence of cosmic birefringence, great care must therefore be taken with spectral 
distortions that may be induced by polarized mixing. 

\section{Conclusion}
\label{sec:conclusion}

In the present work, we discuss how the combination of multiple polarized signals with different spectral parameters and polarization angles (referred to as polarized mixing) unavoidably leads to a different spectral behavior for the polarized angular spectra $EE$, $BB$, and $EB$, thus implying spectral dependence of the $EE/BB$ ratio and nontrivial distortions of the $EB$ correlation. We show how this phenomenon can be understood and tackled using the spin-moment expansion, formally deriving all the analytical expressions at order one in the case of Galactic dust modified black bodies with varying spectral indices while keeping in mind that this would be straightforward when generalizing for any polarized SED (e.g., synchrotron) and at any order. 

We thoroughly discuss the toy model example of a dust-emitting filament in front of a background. A careful understanding of the geometrical and spectral properties of the signal in the filament itself in pixel space allowed us to explain the shape of the total polarization angular power spectra when changing the value of the polarization angle of the filament, $\psifl$, as well as the amplitude of the observed distortions between $100$ and $400$\,GHz. Moreover, we show how one could accurately recover  the spectral dependence of the polarization angular power spectra from the spin-moment maps, validating our previous theoretical considerations. Finally, we considered some of the {\sc PySM} models on the sphere. We show that these models intrinsically contain variations of the $EE/BB$ ratio through the frequency and distortions of the $EB$ correlation, whose amplitudes are expected to strongly depend on the sky fraction and multipole range considered. This allowed us to stress that seeking  a spectral dependence of $EE/BB$ provides a way to explore the existence of polarized mixing and thus independently of the canonical SED used to model the signal. In these {\sc PySM} models, simple assumptions about the frequency dependence of the dust $EB$ signal used in CMB cosmic birefringence analysis become invalid due to the polarized mixing. 

Further studies need to be done in order to precisely assess the expected amplitude of both these effects on real sky data. Meanwhile, it will be necessary to quantify the impact of the assumptions made on the dust $EE/BB$ ratio and $EB$ SEDs on cosmology. Map-based component separation is sensitive to the variation of the foreground's polarization angle with frequency, source of the power-spectrum effects discussed in the present work. However, current $B$-mode only analyses at the power spectrum level are, in principle, immune to the potential variation of the dust $EE/BB$, as they model the $BB$ SED independently of the $EE$ SED. Still, next-generation experiments probing the CMB reionization  bump (as e.g., the $LiteBIRD$ mission \cite{Ptep}) might consider simultaneously $EE$ and $BB$ in order to tackle the correlations between the tensor-to-scalar ratio $r$ and the re-ionization optical depth $\tau$. These experiments would therefore be sensitive to the assumptions made about the dust $EE/BB$ ratio frequency dependence. Regarding the dust $EB$ correlation, as already stressed in \citet{Diego-Palazuelos2022b}, distortions of the MBB SED could impact cosmic birefringence studies to some degree, but as long as this signal remains undetected, using higher signal-to-noise spectra as proxies to $EB$ has to be done with even more caution. In any case, the 3D variation of the dust composition, physical conditions, and the orientation of the Galactic magnetic field produce complex polarization effects in map space and at the angular power-spectrum level, and their impact have to either be ruled out or taken into account for precise $B$-mode and birefringence measurements.
\begin{acknowledgements}
L.V. would like to thanks Aditya Rotti and Louise Mousset for precious feedback and discussions. 
A.R. acknowledges financial support from the Italian Ministry of University and Research - Project Proposal CIR01\_00010. 
JC was supported by an ERC Consolidator Grant {\it CMBSPEC} (No.~725456) and the Royal Society as a Royal Society University Research Fellow at the University of Manchester, UK (No.~URF/R/191023).
\end{acknowledgements}
\bibliographystyle{aa}
\bibliography{bi}

\appendix

\section{Impact of the background polarization angle \label{sec:Appendix-psifg}}

We present the dependence in $\psifl$ of the polarization power spectra of the toy model filament. We consider two different values for the background angle, $\psibg=22.5^{\circ}$ and $\psibg=45^{\circ}$.
\begin{figure}[!ht]
    \centering
    \includegraphics[scale=0.5]{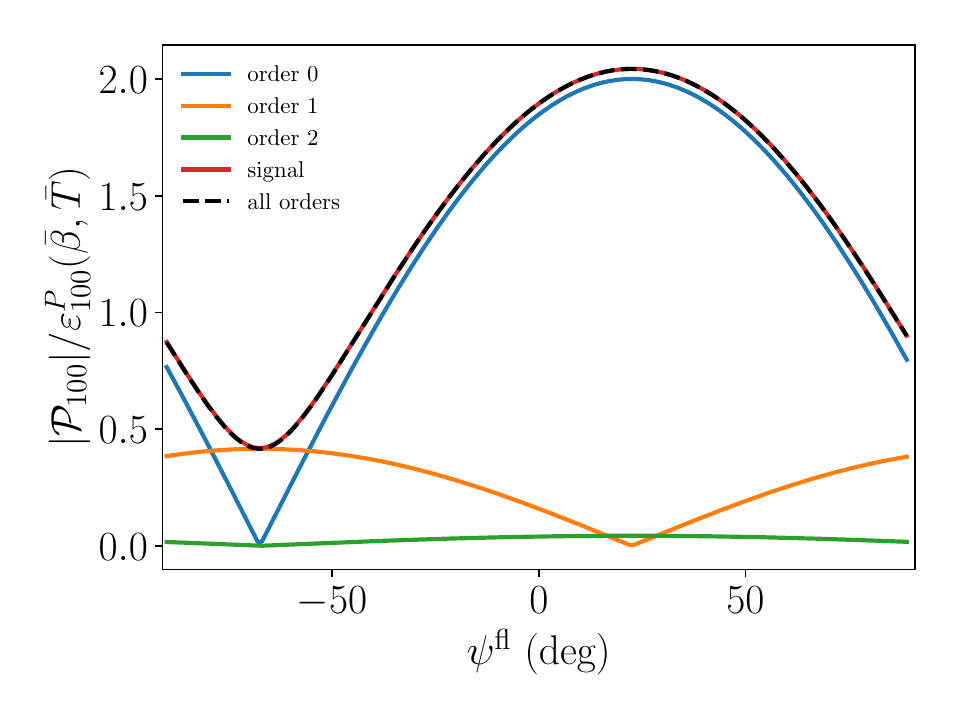}
    \includegraphics[scale=0.5]{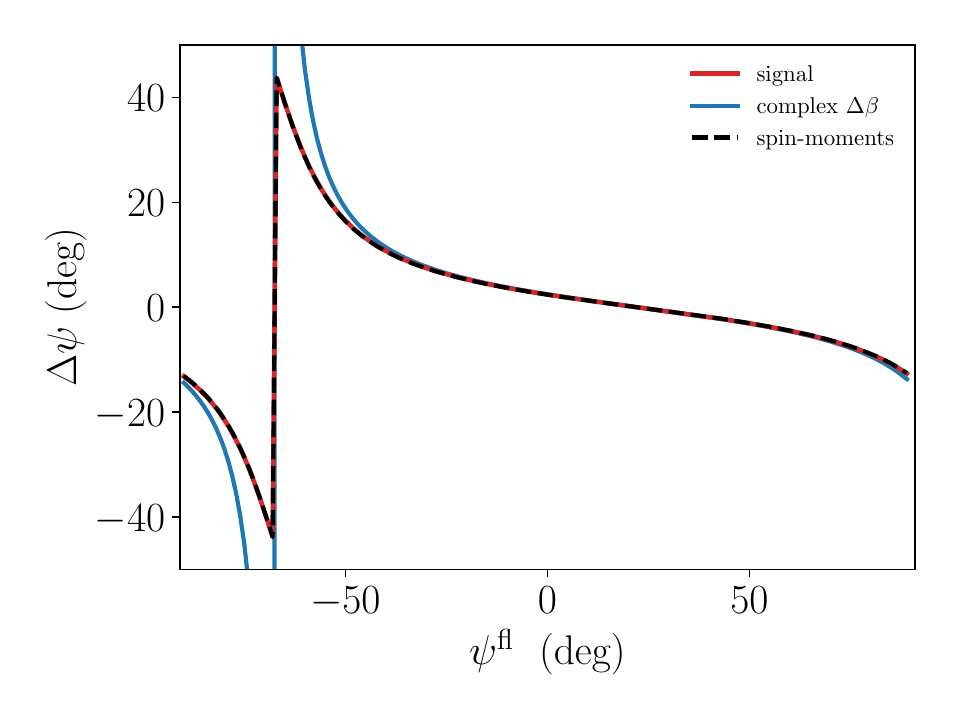}
    \caption{Total polarization spinor in the filament of the toy model for $\psibg=22.5^\circ$ and various values of the filament polarization angle $\psifl$. Left: Modulus of the total signal at $100$\,GHz normalized by the pivot MBB (red) and modulus of the analytical derivation from the spin-moment expansion up to second order (black dashed line). The modulus of each term is displayed: order 0: $|\mathcal{W}_0|$ (blue), order 1: $|\mathcal{W}^\beta_1\ln\left(100/400\right)|$ (orange), and  order 2: $|0.5\mathcal{W}^\beta_2\ln\left(100/400\right)^2|$ (green).  Right: Difference of the polarization angles between the two frequencies (red), prediction from the complex $\Delta\beta$ correction $0.5\,{\rm Im}(\mathcal{W}^\beta_1/\mathcal{W}_0)\ln\left(100/400\right)$ (blue) and from the spin-moment expansion up to second order (black dashed line). }
\label{fig:signalpsibg22.5}
\end{figure}
\begin{figure}[!ht]
    \centering
    \includegraphics[scale=0.5]{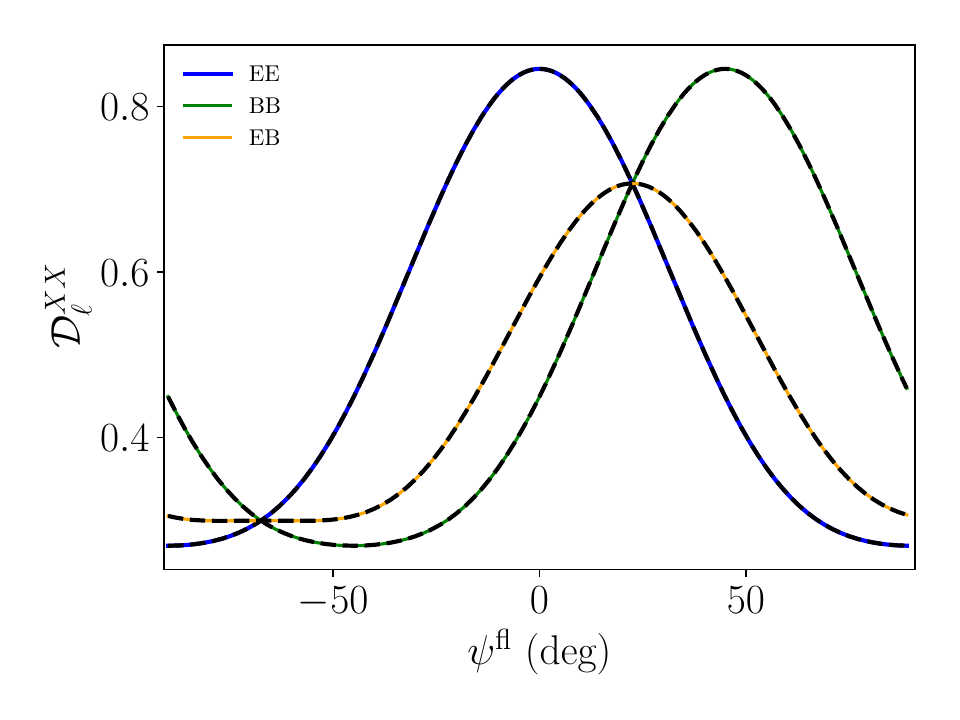}
    \includegraphics[scale=0.5]{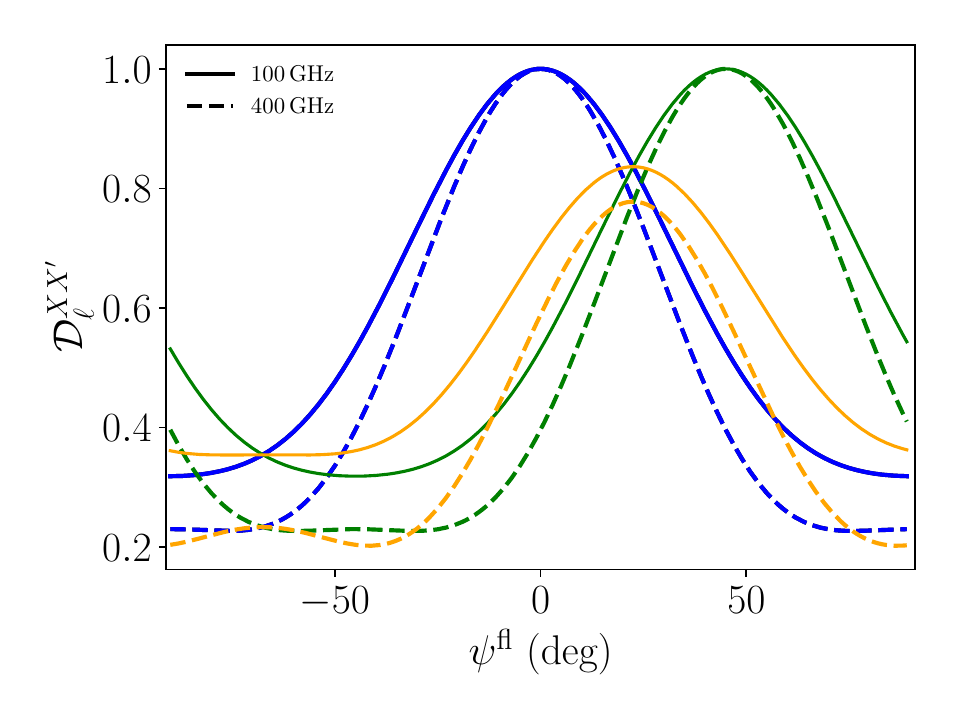}
    \caption{Mean polarized power spectra for the toy model filament with $\psibg=22.5^\circ$ and various values of the filament polarization angle $\psifl$. The $EE$ (blue), $BB$ (green), and $EB$ (orange) angular power spectra are given at two different frequencies: 100 (continuous) and 400 GHz (dashed). Each spectra is normalized by the maximum value of $\left(\left(\DLEE\right)^2(\nu) + \left(\DLBB\right)^2(\nu)\right)^{1/2}$. Left: No SED distortions: $\beta^{\rm fl}=\beta^{\rm bg}=1.5$. Right: with SED distortions: $\beta^{\rm fl}=1.8$ and $\beta^{\rm bg}=1.5$.}
\label{fig:EEBBpsibg22.5}
\end{figure}
\begin{figure}[!ht]
    \centering
    \includegraphics[scale=0.5]{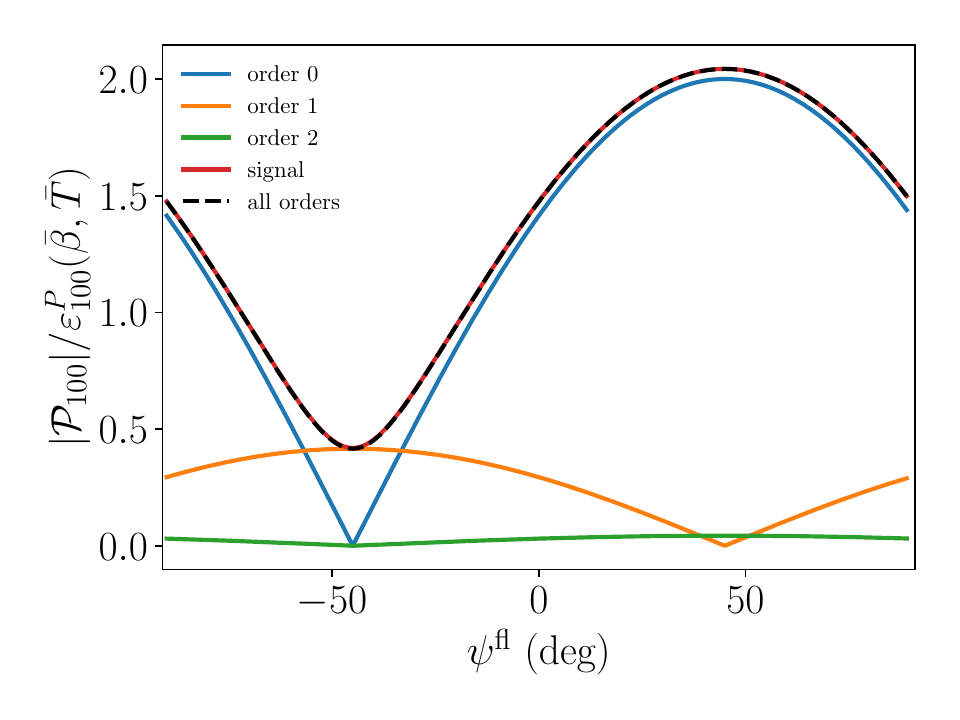}
    \includegraphics[scale=0.5]{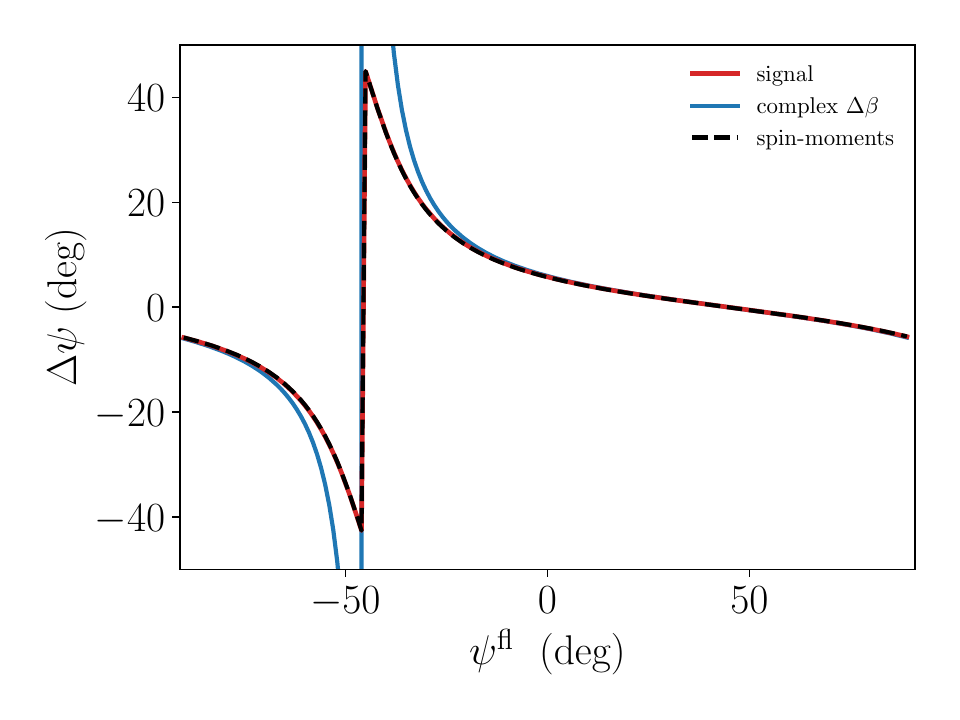}
    \caption{Same as Fig.~\ref{fig:signalpsibg22.5} but for $\psibg=45^\circ$.}
\end{figure}
\begin{figure}[!ht]
    \centering
    \includegraphics[scale=0.5]{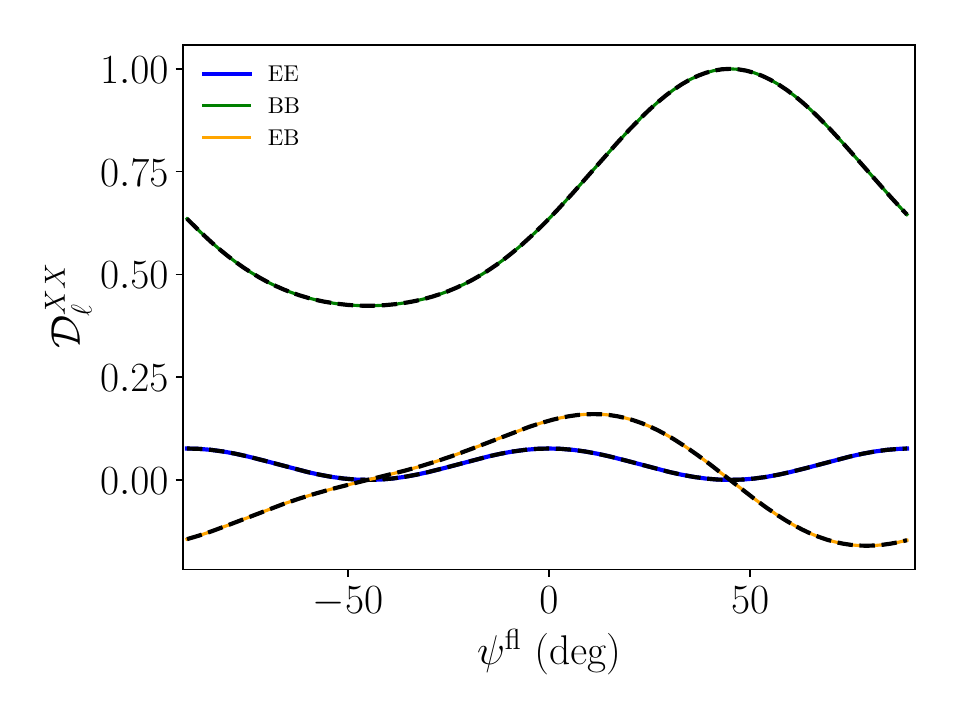}
    \includegraphics[scale=0.5]{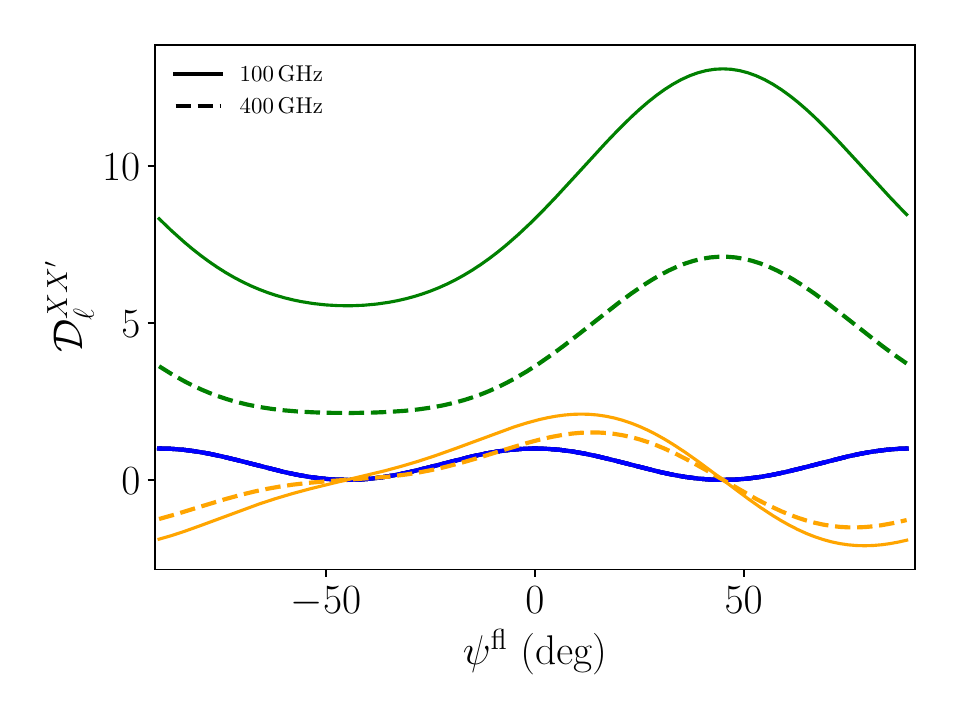}

    \caption{Same as Fig.~\ref{fig:EEBBpsibg22.5} but for $\psibg=45^\circ$.}
\end{figure}

\end{document}